\documentclass[superscriptaddress,prd,aps,showpacs,nofootinbib,showkeywords,eqsecnum]{revtex4-2}

\usepackage{graphicx,color,amsmath,amsxtra}
\usepackage{epsf}
\usepackage{amssymb}
\usepackage{enumerate}
\usepackage{hhline}
\usepackage{array}
\usepackage{tabularx}
\usepackage[unicode]{hyperref}              

\begin{document}
\pagestyle{myheadings}
\title{Unstable de Sitter inflationary solution in sixth-order gravity}
\author{Tuan Q. Do}
\email{tuan.doquoc@phenikaa-uni.edu.vn}
\affiliation{Phenikaa Institute for Advanced Study, Phenikaa University, Hanoi 12116, Vietnam}

\begin{abstract}
A sixth-order gravity, which involves not only two leading terms, $\gamma_1 R\Box R$ and $\gamma_2 R_{\mu\nu} \Box R^{\mu\nu}$, but also quadratic curvature terms along with cubic curvature ones, will be investigated in this paper to see if it admits an exact stable de Sitter solution.  First, we will derive sixth-order differential field equations of this gravity under the homogeneous and isotropic Friedmann-Lemaitre-Robertson-Walker background spacetime, using the effective method based on the Euler-Lagrange equations. Then, we will analytically solve these field equations to figure out an exact de Sitter solution, which turns out to be equivalent to a fixed point of the corresponding dynamical system of the studied gravity. Interestingly, two coefficients, $\gamma_1$ and $\gamma_2$, do not contribute to the value of the obtained de Sitter solution. However, they affect on the stability of the de Sitter solution. In particular, if these two coefficients obey the following inequality, $3\gamma_1+\gamma_2 <0$, then the de Sitter solution will always be unstable. Furthermore, numerical calculations will be performed to verify that the de Sitter fixed point is indeed a repeller of the dynamical system once this inequality is satisfied. All these results indicate that the sixth-order gravity is more suitable for an  inflationary phase of early universe. To be complete, two special limits of the studied gravity model, in which field equations are reduced to second-order and fourth-order, respectively, will be investigated. As expected, only the second-order limit can always give raise a stable de Sitter solution, compatible with an accelerated expansion of  late-time universe. 
\end{abstract}
\maketitle
\newpage
\section{Introduction} \label{intro}
An inflationary universe paradigm has led the rapid development progress of theoretical cosmology over four recent decades since the pioneered works of Starobinsky, Guth, Linde, and many others \cite{Starobinsky:1980te,Guth:1980zm,Linde:1981mu,Linde:1983gd,Baumann}. Remarkably, it has provided a theoretical framework to guide the cosmic microwave background radiations probes such as the Planck satellite \cite{Akrami:2018odb} and the Atacama Cosmology Telescope (ACT)\cite{AtacamaCosmologyTelescope:2025nti}. Basically, there have been two main directions in realizing the mechanism of generating the cosmic inflation of early universe. The first one is due to an introduction of additional scalar field called an inflaton, whose very first examples were proposed in Refs. \cite{Guth:1980zm,Linde:1981mu,Linde:1983gd}, while the last one is purely geometrical, whose very first example was introduced in Ref. \cite{Starobinsky:1980te}. It is worth noting that the latter one, which has been widely regarded as the Starobinsky model, has remained as one of the most viable inflationary models in the light of the latest  Planck data \cite{Akrami:2018odb}. 

However, recent ACT data \cite{AtacamaCosmologyTelescope:2025nti} has pushed the Starobinsky model into a serious tension. In particular, the value of scalar spectral index, $n_s$, observed by the ACT is slightly higher than the theoretical value derived in the Starobinsky model. Since then, the Starobinsky model has been re-considered extensively by many people, e.g., see Refs. \cite{Addazi:2025qra,Bianchi:2025tyl,Ketov:2025cqg,Ellis:2025ieh,Park:2025upd,Morais:2025ski,Bezerra-Sobrinho:2025gfg} for some interesting examples. It turns out that most of these considerations are extending the Starobinsky model by introducing higher-order curvature corrections into its action and recalculating the corresponding scalar spectral index, $n_s$, as well as tensor-to-scalar ratio, $r$. Theoretically, going beyond the Starobinsky model was actively considered in the past, even when the ACT data was not officially announced, e.g., see Refs. \cite{Whitt:1984pd,Maeda:1987xf,Barrow:1988xh,Faraoni:2004dn,Faraoni:2005vk,Faraoni:2007yn,Barrow:2006xb,Toporensky:2006kc,Huang:2013hsb,Elizalde:2014xva,Pozdeeva:2019agu,Vernov:2021hxo,Myrzakulov:2014hca,Aoki:2019snr,Cheong:2020rao,Cano:2020oaa,Rodrigues-da-Silva:2021jab,Ivanov:2021chn,Ketov:2022lhx,Shtanov:2022pdx,Modak:2022gol,Ketov:2022zhp,Asorey:2024oxw,Lambiase:2025qyl} for some interesting works, which are currently relevant to our studies on higher-order gravities. 

Mathematically, the Starobinsky model can be classified as a leading subclass of the fourth-order gravity, whose cosmological consequences for the early universe are very rich \cite{Schmidt:2006jt,Alvarez-Gaume:2015rwa,Salvio:2018crh}. It can also be classified as the simplest subclass of the so-called $f(R)$ gravity, which has been expected to provide a resolution to an accelerated expansion of  late-time universe \cite{Copeland:2006wr,Sotiriou:2008rp,DeFelice:2010aj,Carroll:2004de,Amendola:2006we,Nojiri:2010wj,Nojiri:2017ncd,Capozziello:2011et}. Besides, the Starobinsky model could be relevant to quantum gravity as claimed in Ref. \cite{Stelle:1976gc}. For this issue, one may want to see a recent interesting work on the so-called Batalin-Fradkin-Vilkovisky quantization of quadratic gravity \cite{Bellorin:2025rtw}.  Very interestingly, although it is the fourth-order gravity but it is free from the so-called Ostrogradsky ghost \cite{Woodard:2015zca}.  Recently, it has shown in Ref. \cite{cubic} that the $R^2$ term, which is key ingredient of the Starobinsky model as well as the quadratic gravity \cite{Alvarez-Gaume:2015rwa,Salvio:2018crh}, can help to resolve a serious issue on the stability of de Sitter solution of the generalized Einsteinian cubic gravity \cite{DeFelice:2023vmj}.  All of these properties support the uniqueness of the Starobinsky model \cite{Ketov:2025nkr}. 

So far, almost all non-trivial extensions of the Starobinsky model have been constructed to admit fourth-order field equations in the FLRW background, with the expectation that they do not imply stable de Sitter inflationary solutions \cite{Ketov:2025nkr}.  One can therefore ask if other corrections, which lead to the existence of higher-than-fourth-order terms in field equations, could still generate unstable de Sitter inflationary solutions.  In this paper, we would like to address this important question by examining whether the so-called sixth-order gravity \cite{Giacchini:2025gzw,Giacchini:2024exc} admits a stable de Sitter solution as its cosmological solution or not. This examination is an important criterion to determine which phase of universe would be compatible with the sixth-order gravity. Historically,  sixth-order gravities were proposed and investigated quite a long time ago \cite{Xu:1987pz,Gottlober:1989ww,Schmidt:1990dh,Berkin:1990nu,Quandt:1990gc,Fulling:1992vm,Amendola:1993bg,Harvey,Decanini:2007gj,Skugoreva:2010dm,Iihoshi:2010pf}. It is worth noting that cosmic inflation was discussed in some simple sixth-order gravities in Refs. \cite{Gottlober:1989ww,Schmidt:1990dh,Berkin:1990nu,Amendola:1993bg}. For example, unstable de Sitter and power law inflationary solutions were found in Ref. \cite{Berkin:1990nu,Amendola:1993bg}, while quasi-de Sitter ones were pointed out to exist in Ref. \cite{Gottlober:1989ww}.  More interestingly, sixth-order gravities  can be conformally transformed into effective theories of two scalar fields as proved in  Refs. \cite{Gottlober:1989ww,Schmidt:1990dh}.  The sixth-order gravity considered in Refs. \cite{Giacchini:2025gzw,Giacchini:2024exc} seems to be more complicated since it includes not only $R \Box R$ and $R_{\mu\nu} \Box R^{\mu\nu}$, which merely generate sixth-order derivatives in field equations, but also cubic curvature terms. Very interestingly, this gravity still involves quadratic curvature terms, i.e., $R^2$ and $R_{\mu\nu}R^{\mu\nu}$. Therefore, the six-order gravity proposed in Refs. \cite{Giacchini:2025gzw,Giacchini:2024exc} can be regarded as an ``exotic" extension of the Starobinsky model. Remarkably,  the typical term of the sixth-order gravity, i.e., $R \Box R$, has been shown to contribute only a correction to the Starobinsky model during the inflationary phase as shown in a recent paper \cite{Khodabakhshi:2024med}.

Then, we will examine the effect of the quadratic curvature terms, $R^2$ and $R_{\mu\nu}R^{\mu\nu}$, on the stability of the obtained de Sitter solution, similar to our previous study in Ref. \cite{cubic}. Specifically, we will  turn off the $R^2$ and $R_{\mu\nu}R^{\mu\nu}$ terms to see if the corresponding set of perturbed equations is incomplete. In harmony with our recent paper \cite{cubic}, we will consider two special limits, one is the fourth-order limit and the other is the second-order limit, to see whether their corresponding de Sitter solutions are stable or not. It should be noted that all theoretical calculation techniques used in the present paper have been successfully implemented in our recent papers \cite{cubic,Do:2020vdc,Do:2021fal,Do:2023yvg,Pham:2024fub,Do:2026pkg,preprint}.

Our paper will be organized as follows: (i) Section \ref{intro} has been used to present a brief introduction of our study. (ii) Section \ref{sec2} will be devoted to present the basic setup of the sixth-order gravity under the Friedmann-Lemaitre-Robertson-Walker (FLRW) background, including its general field equations and two special limits, in which the field equations are only either second-order or fourth-order ordinary differential equations (ODEs). (iii) In Section \ref{sec3}, we will figure out an exact de Sitter solutions to the field equations and analyse their stability via the dynamical system approach not only in the sixth-order case but also in second-order and fourth-order limits. Through this analysis, we will be able to examine the effect of the quadratic curvature terms on the stability of the obtained de Sitter solutions. (iv) Finally, Section \ref{final} will be our concluding remarks. It is noted that  additional calculations will be presented in the Appendix \ref{app} just for ensuring the validity of the obtained results in the main text. 
\section{Sixth-order gravity} \label{sec2}
\subsection{Action}
We would like to consider the following action of sixth-order gravity \cite{Giacchini:2024exc,Giacchini:2025gzw}
\begin{align} \label{action-1}
S  = & \int d^4 x \sqrt{-g} \left[   \alpha  R  
+ \beta_1 R^2 +  \beta_2 R_{\mu\nu}^2 
+ \gamma_1 R \Box R 
+ \gamma_2 R_{\mu\nu} \Box R^{\mu\nu} 
+ \gamma_3 R^3 
+ \gamma_4 R R_{\mu\nu} R^{\mu\nu} \right. \nonumber\\
& \left.  +  \gamma_5 R_{\mu\nu} R^\mu{}_\rho R^{\rho \nu}
+ \gamma_6 R_{\mu\nu} R_{\rho\sigma} R^{\mu\rho\nu\sigma} 
+ \gamma_7 R R_{\mu\nu\rho\sigma} R^{\mu\nu\rho\sigma}
+ \, \gamma_8 R_{\mu\nu\rho\sigma} R^{\mu\nu\tau\upsilon} R^{\rho\sigma}{}_{\tau\upsilon} 
\right],
\end{align}
where $\alpha$, $\beta_{i}~(i=1,~2)$, and $\gamma_{j}~(j=1,~2,~3,~4,~5,~6,~7,~8)$ are the coefficients of the terms with a total of two, four, and six metric derivatives, respectively. In the above action, the well-known factor $1/(16\pi G)$  has been absorbed into the coefficients  just for convenience. As stated in Refs. \cite{Giacchini:2024exc,Giacchini:2025gzw}, this action is the most general extension of the Einstein-Hilbert action by terms with a total of four and six derivatives of the metric. Indeed, all  other possible terms, e.g., see those in Ref. \cite{Fulling:1992vm}, can be formed as combinations of the terms shown in the action \eqref{action-1} and boundary or topological terms, which do not contribute to the dynamics of model, according to Refs. \cite{Xu:1987pz,Harvey,Decanini:2007gj}. In the action \eqref{action-1}, $\Box$ is the d'Alembertian operator, which acts on $R$ and $R^{\mu\nu}$ as follows
\begin{equation}
\Box R \equiv \nabla^\nu (\nabla_\nu R), \quad \Box R^{\mu\nu} \equiv \nabla^\rho (\nabla_\rho R^{\mu\nu}).
\end{equation}
It is worth noting that  simpler versions of sixth-order gravity were considered quite a long time ago \cite{Gottlober:1989ww,Schmidt:1990dh,Berkin:1990nu,Quandt:1990gc,Amendola:1993bg}. See also Refs. \cite{Skugoreva:2010dm,Iihoshi:2010pf,Khodabakhshi:2024med} for recent considerations. As will be shown below for the FLRW metric, sixth-order derivatives in field equations will only be generated from $\gamma_1$- and $\gamma_2$-terms involving $R \Box R$ and $R_{\mu\nu}\Box R^{\mu\nu}$. The other $\gamma_j$-terms $(j=3,~4,~5,~6,~7,~8)$ are nothing but that appearing in the cubic gravities \cite{Bueno:2016xff,Hennigar:2017ego,DeFelice:2023vmj,Arciniega:2018fxj,Arciniega:2018tnn,Pookkillath:2020iqq}, which give rise fourth-order derivatives in field equations. Before ending this subsection, it is important to note that a $n$-th--order gravity can be obtained by introducing a term $R \Box^{k} R$ into the Einstein-Hilbert action, where $n=2k+4 \geq 8$ with $k\in \mathbb{Z}^+$ and $k\geq 2$, according to Ref. \cite{BattagliaMayer:1993yf}.
\subsection{Field equations for the FLRW background}
Since we are seeking de Sitter solutions to this gravity, we will consider the spatially flat $(k=0)$ FLRW metric given by \cite{cubic}
\begin{equation} \label{metric}
ds^2 =-N^2(t)dt^2 +e^{2\beta(t)} \left(dx^2 + dy^2 +dz^2 \right).
\end{equation}
It is important to note that the FLRW metric in a unique spacetime obeying the underlying assumption of modern cosmology, which is called the cosmological principle stating that our universe is simply homogeneous and isotropic on large scales. These properties can be seen from the scale factor characterized by $\beta(t)$, which depends only on the cosmic time $t$. In the above expression, we have introduced $N(t)$, which is nothing but the lapse function of $t$. It is apparent that the existence of $N(t)$ is vital to derive a full set of field equations using the well-known technique based on the Euler-Lagrange (EL) equations.  This technique has turned out to be an effective method to derive the field equations of higher-order gravities, whose tensorial field equations (a.k.a. Einstein field equations) are usually complicated to handle. Indeed, we have used this method in a number of published papers, e.g., Refs. \cite{cubic,Do:2020vdc,Do:2021fal,Do:2023yvg,Pham:2024fub,Do:2026pkg,preprint}. Interestingly, other people have also used this method for their studies as well, e.g., see Ref. \cite{Asorey:2024oxw} for a recent related study. 

First, we are going to determine the corresponding Lagrangian of the  sixth-order gravity, whose general expression is given by
\begin{align}
{\cal L} = &~ \sqrt{-g} \left[   \alpha  R  
+ \beta_1 R^2 +  \beta_2 R_{\mu\nu}^2 
+ \gamma_1 R \Box R 
+ \gamma_2 R_{\mu\nu} \Box R^{\mu\nu} 
+ \gamma_3 R^3 
+ \gamma_4 R R_{\mu\nu} R^{\mu\nu} \right. \nonumber\\
& \left.  +  \gamma_5 R_{\mu\nu} R^\mu{}_\rho R^{\rho \nu}
+ \gamma_6 R_{\mu\nu} R_{\rho\sigma} R^{\mu\rho\nu\sigma} 
+ \gamma_7 R R_{\mu\nu\rho\sigma} R^{\mu\nu\rho\sigma}
+ \, \gamma_8 R_{\mu\nu\rho\sigma} R^{\mu\nu\tau\upsilon} R^{\rho\sigma}{}_{\tau\upsilon} 
\right].
\end{align}
It is useful to work out explicit expressions of all terms appearing in this Lagrangian. As a result, we are able to obtain the following results  \cite{xAct},
\begin{align}
\sqrt{-g} = &~ N e^{3\beta},\\
R=&~ \frac{6}{N^3} \left[ N \left(\ddot\beta+2 \dot\beta^2\right)- \dot N \dot\beta \right],\\
R_{\mu\nu}^2 \equiv R_{\mu\nu}R^{\mu\nu} = &~\frac{12}{N^6} \left[ \dot N^2 \dot\beta^2-N \dot N \dot\beta \left(2 \ddot\beta+3 \dot\beta^2\right)+N^2 \left(\ddot\beta^2+3 \dot\beta^2 \ddot\beta+3 \dot\beta^4\right) \right],\\
R_{\mu\nu} R^\mu{}_\rho R^{\rho \nu}=&~\frac{3}{N^9} \left\{ \left[ N \left(\ddot\beta+3 \dot\beta^2\right)-\dot N \dot\beta\right]^3-9 \left[ \dot N \dot\beta-N \left(\ddot\beta+\dot\beta^2\right)\right] ^3 \right\},\\
R_{\mu\nu} R_{\rho\sigma} R^{\mu\rho\nu\sigma}=&~0,\\
R R_{\mu\nu\rho\sigma} R^{\mu\nu\rho\sigma} =& - \frac{72}{N^9} \left[ \dot N^3 \dot\beta^3+ N^2 \dot N \dot\beta \left(3 \ddot\beta^2+8 \dot\beta^2 \ddot\beta+6 \dot\beta^4\right) \right. \nonumber\\
&\left. - N \dot N^2 \dot\beta^2 \left(3 \ddot\beta+4 \dot\beta^2\right)-N^3 \left(\ddot\beta^3 +4 \dot\beta^2 \ddot\beta^2 +6 \dot\beta^4 \ddot\beta +4 \dot\beta^6\right)\right],\\
R_{\mu\nu\rho\sigma} R^{\mu\nu\tau\upsilon} R^{\rho\sigma}{}_{\tau\upsilon} =&-\frac{24}{N^9} \left[ \dot N^3 \dot\beta^3+3 N^2 \dot N \dot\beta \left(\ddot\beta+\dot\beta^2\right)^2-3 N \dot N^2 \dot\beta^2 \left(\ddot\beta+\dot\beta^2\right) \right. \nonumber\\
& \left. -N^3 \left(\ddot\beta^3+3 \dot\beta^2 \ddot\beta^2+3 \dot\beta^4 \ddot\beta +2 \dot\beta^6 \right)\right],\\
\Box R =&~ \frac{6}{N^7} \left\{ 15 \dot N^3 \dot\beta-5 N \dot N^2 \left(3 \ddot\beta+5 \dot\beta^2\right) \right. \nonumber\\
& \left.  +N^2 \left[  \dot\beta \left(N^{(3)}-7 N \beta^{(3)}\right)+4 \ddot N \ddot\beta+\dot\beta^2 \left(7 \ddot N-12 N \ddot\beta\right) \right. \right. \nonumber\\
& \left. \left. -N \left(\beta^{(4)}+4 \ddot\beta^2\right)\right]+N \dot N \left[ N \left(6 \beta^{(3)}+29 \dot\beta \ddot\beta+12 \dot\beta^3\right)-10 \ddot N \dot\beta\right]\right\},\\
R_{\mu\nu} \Box R^{\mu\nu} =&~-\frac{6}{N^{10}} \left\{ 30 \dot N^4 \dot\beta^2-3 N \dot N^3 \dot\beta \left(20 \ddot\beta+29 \dot\beta^2\right) \right. \nonumber\\
&\left. -N^3 \left[ 3\dot\beta^3 \left(N^{(3)}-7 N \beta^{(3)}\right) + 2\dot\beta \ddot\beta  \left(N^{(3)}-6 N \beta^{(3)}\right)  \right. \right. \nonumber\\
& \left. \left. + 2 \ddot\beta \left( 4 \ddot N \ddot\beta - N \left(\beta^{(4)}+3 \ddot\beta^2\right) \right)+ \dot\beta^4 \left(21 \ddot N-36 N \ddot\beta\right)  \right.\right. \nonumber\\
&\left. \left. + \dot\beta^2 \left( 24 \ddot N \ddot\beta -N \left(3 \beta^{(4)}+26 \ddot\beta^2\right)\right)\right] \right. \nonumber\\
&\left. - N \dot N^2 \left[ 20 \ddot N \dot\beta^2-N \left(12\dot\beta \beta^{(3)} +30 \ddot\beta^2+135 \dot\beta^2 \ddot\beta +89 \dot\beta^4 \right)\right] \right. \nonumber\\
&\left. + N^2 \dot N \left[ 2 \dot\beta \left(N^{(3)} \dot\beta+14 \ddot N \ddot\beta+21 \ddot N \dot\beta^2\right) \right. \right. \nonumber\\
&\left. \left. -N \left(2 \dot\beta \beta^{(4)}+12\ddot\beta \beta^{(3)} +30  \dot\beta^2 \beta^{(3)}+54 \dot\beta \ddot\beta^2+115 \dot\beta^3 \ddot\beta +36 \dot\beta^5 \right) \right] \right\}.
\end{align}
Here, the higher-order time derivatives have been denoted such as $\beta^{(n)} \equiv d^n \beta/ dt^n$ with $n \geq 3$. On the other hand, the first- and second-order time derivatives follow the usual notations, $\dot\beta \equiv d\beta/dt$ and $\ddot\beta \equiv d^2 \beta/dt^2$, respectively.  
The vanishing of the $\gamma_6$-term is simply due to the antisymmetric property of Riemann tensor. 
Since ${\cal L}$ is a functional of third-order time derivative of $N$, the EL equation for $N$ takes the following form,  
\begin{equation}
\frac{\partial {\cal L}}{\partial N} -\frac{d}{dt} \left(\frac{\partial {\cal L}}{\partial \dot N}\right) +\frac{d^2}{dt^2} \left(\frac{\partial {\cal L}}{\partial \ddot N}\right) - \frac{d^3}{dt^3} \left(\frac{\partial {\cal L}}{\partial N^{(3)}}\right) =0.
\end{equation}
On the other hand, the EL equation for $\beta$ reads
\begin{equation}
\frac{\partial {\cal L}}{\partial \beta} -\frac{d}{dt} \left(\frac{\partial {\cal L}}{\partial \dot \beta}\right) + \frac{d^2}{dt^2} \left(\frac{\partial {\cal L}}{\partial \ddot\beta}\right)-\frac{d^3}{dt^3} \left(\frac{\partial {\cal L}}{\partial \beta^{(3)}}\right)+\frac{d^4}{dt^4} \left(\frac{\partial {\cal L}}{\partial \beta^{(4)}}\right)=0,
\end{equation}
due to the fact that ${\cal L}$ is a  functional of fourth-order time derivative of $\beta$.
Thanks to the results shown above, the explicit expression of the first EL equation is defined to be
\begin{align} \label{first-field-equation}
& \alpha \dot\beta^2 + 2\left( 3 \beta_1 +\beta_2 \right)  \left(2\dot\beta  \beta ^{(3)}  -{\ddot\beta}^2+6 {\dot\beta}^2 \ddot\beta \right) \nonumber\\
& -6 \gamma_1  \left( 2\dot\beta \beta ^{(5)}-2\ddot\beta \beta ^{(4)}   +12{\dot\beta}^2 \beta ^{(4)} + {\beta ^{(3)}}^2+24\dot\beta \ddot\beta \beta ^{(3)} +10{\dot\beta}^3 \beta ^{(3)}  -8 {\ddot\beta}^3+32 {\dot\beta}^2 {\ddot\beta}^2 -24 {\dot\beta}^4 \ddot\beta \right) \nonumber\\
& - 2\gamma_2  \left( 2 \dot\beta \beta ^{(5)} -2  \ddot\beta \beta ^{(4)}+12{\dot\beta}^2 \beta ^{(4)} + {\beta ^{(3)}}^2 +18\dot\beta \ddot\beta \beta ^{(3)} +8 {\dot\beta}^3 \beta ^{(3)}  - 6 {\ddot\beta}^3 +22 {\dot\beta}^2 {\ddot\beta}^2-30 {\dot\beta}^4 \ddot\beta \right) \nonumber\\
& +36 \gamma_3  \left(\ddot\beta+2 {\dot\beta}^2\right) \left(6  \dot\beta \beta ^{(3)} -2 {\ddot\beta}^2+19 {\dot\beta}^2 \ddot\beta-2 {\dot\beta}^4\right) \nonumber\\
& +12\gamma_4 \left(6  \dot\beta \ddot\beta \beta ^{(3)} +10{\dot\beta}^3 \beta ^{(3)}-2 {\ddot\beta}^3+14 {\dot\beta}^2 {\ddot\beta}^2 +30 {\dot\beta}^4 \ddot\beta -3 {\dot\beta}^6 \right) \nonumber\\
& +\gamma_5 \left(30\dot\beta \ddot\beta \beta ^{(3)} +36{\dot\beta}^3 \beta ^{(3)}  -10 {\ddot\beta}^3+63 {\dot\beta}^2 {\ddot\beta}^2 +108 {\dot\beta}^4 \ddot\beta -9 {\dot\beta}^6\right) \nonumber\\
&+12\gamma_7 \left(6\dot\beta \ddot\beta \beta ^{(3)} +8 {\dot\beta}^3 \beta ^{(3)} -2 {\ddot\beta}^3+13 {\dot\beta}^2 {\ddot\beta}^2 +24 {\dot\beta}^4 \ddot\beta -2 {\dot\beta}^6 \right) \nonumber\\
& +4\gamma_8 \left( 6\dot\beta \ddot\beta \beta ^{(3)}+ 6 {\dot\beta}^3 \beta ^{(3)} -2 {\ddot\beta}^3+12 {\dot\beta}^2 {\ddot\beta}^2  +18 {\dot\beta}^4 \ddot\beta - {\dot\beta}^6 \right) =0.
\end{align}
On the other hand, the second EL equation explicitly becomes  as 
\begin{align} \label{second-field-equation}
&\alpha \left(2 \ddot\beta+3 {\dot\beta}^2 \right) + 2\left(3 \beta_1 +\beta_2 \right)  \left(2 \beta ^{(4)}+12\dot\beta \beta ^{(3)} +9 {\ddot\beta}^2+18 {\dot\beta}^2 \ddot\beta\right) \nonumber\\
& - 6\gamma_1 \left(2 \beta ^{(6)}+18\dot\beta  \beta ^{(5)} +42\ddot\beta \beta ^{(4)} +46{\dot\beta}^2  \beta ^{(4)}  +27 {\beta ^{(3)}}^2   +166 \dot\beta \ddot\beta \beta ^{(3)}  +6{\dot\beta}^3 \beta ^{(3)} +40 {\ddot\beta}^3 -72 {\dot\beta}^4 \ddot\beta \right) \nonumber\\
& -2\gamma_2  \left(2 \beta ^{(6)}+18\dot\beta  \beta ^{(5)}  +36 \ddot\beta \beta ^{(4)} +44 {\dot\beta}^2 \beta ^{(4)}+21 {\beta ^{(3)}}^2 +122\dot\beta \ddot\beta \beta ^{(3)} -6 {\dot\beta}^3 \beta ^{(3)}  +26 {\ddot\beta}^3 -54 {\dot\beta}^2 {\ddot\beta}^2 -90 {\dot\beta}^4 \ddot\beta \right) \nonumber\\
& +108\gamma_3 \left( 2\ddot\beta  \beta ^{(4)} +4{\dot\beta}^2 \beta ^{(4)} +2 {\beta ^{(3)}}^2+28\dot\beta \ddot\beta  \beta ^{(3)}   +24{\dot\beta}^3 \beta ^{(3)} +8 {\ddot\beta}^3 +63 {\dot\beta}^2 {\ddot\beta}^2    +28 {\dot\beta}^4 \ddot\beta -4 {\dot\beta}^6\right) \nonumber\\
& +12\gamma_4 \left(6\ddot\beta  \beta ^{(4)} +10{\dot\beta}^2 \beta ^{(4)}+ 6 {\beta ^{(3)}}^2 +76\dot\beta \ddot\beta \beta ^{(3)} +60 {\dot\beta}^3  \beta ^{(3)}  +22 {\ddot\beta}^3 +162 {\dot\beta}^2 {\ddot\beta}^2+72 {\dot\beta}^4 \ddot\beta  -9 {\dot\beta}^6  \right) \nonumber\\
& +3\gamma_5 \left(10 \ddot\beta  \beta ^{(4)}+12{\dot\beta}^2 \beta ^{(4)}+10 {\beta ^{(3)}}^2+108\dot\beta \ddot\beta \beta ^{(3)}  +72{\dot\beta}^3 \beta ^{(3)} +32 {\ddot\beta}^3+207 {\dot\beta}^2 {\ddot\beta}^2 +90 {\dot\beta}^4 \ddot\beta -9 {\dot\beta}^6\right) \nonumber\\
& +12 \gamma_7 \left(6\ddot\beta \beta ^{(4)} +8 {\dot\beta}^2 \beta ^{(4)} +6 {\beta ^{(3)}}^2+68\dot\beta \ddot\beta \beta ^{(3)}+48{\dot\beta}^3 \beta ^{(3)}  +20 {\ddot\beta}^3+135 {\dot\beta}^2 {\ddot\beta}^2 +60 {\dot\beta}^4 \ddot\beta -6 {\dot\beta}^6  \right) \nonumber\\
& +12\gamma_8 \left(2\ddot\beta \beta ^{(4)} +2{\dot\beta}^2 \beta ^{(4)}+2 {\beta ^{(3)}}^2+20\dot\beta \ddot\beta \beta ^{(3)} +12{\dot\beta}^3 \beta ^{(3)}   +6 {\ddot\beta}^3+36 {\dot\beta}^2 {\ddot\beta}^2   +16 {\dot\beta}^4 \ddot\beta -{\dot\beta}^6\right)=0.
\end{align}
It is now clear why people call the considered model a sixth-order gravity. The reason is due to the fact that the second EL equation is reduced to a sixth-order ODE of $\beta(t)$ in the FLRW background. It appears that only the $\gamma_1$- and $\gamma_2$-terms introduce the fifth- and sixth-order time derivatives of $\beta(t)$.  On the other hand, the other terms except the Ricci scalar, $R$, all introduce the fourth-order time derivative of $\beta(t)$ as the highest-order derivative. This finding is indeed consistent with the previous studies \cite{Capozziello:2011et,Giacchini:2024exc,Giacchini:2025gzw,Gottlober:1989ww,Schmidt:1990dh,Berkin:1990nu,Quandt:1990gc,Decanini:2007gj,Skugoreva:2010dm}. It is interesting to note that the order of the considered gravity depends on the preferred field equations and/or variables. For example, due to the fact that Eq. \eqref{second-field-equation} can be shown to be a differential consequence of Eq. \eqref{first-field-equation}, which is consistent with the Bianchi identity, one may ignore it for further analysis. Indeed, it is easily shown that
\begin{equation}
{\text{Eq.~}}\eqref{second-field-equation}=3\times {\text{Eq.~}}\eqref{first-field-equation}+\frac{1}{\dot\beta}\times \frac{d}{dt} \left[ {\text{Eq.~}}\eqref{first-field-equation}\right].
\end{equation}
 In this case, therefore, the considered gravity model may be called a fifth-order gravity rather than a sixth-order one because Eq. \eqref{first-field-equation} is just the fifth-order ODE. On the other hand, by introducing the Hubble parameter as $H \equiv \dot\beta$, the order of the field equations can be reduced from the sixth order to the fifth order and so on. An interesting example can be seen in Ref. \cite{Toporensky:2006kc}. See also our recent paper \cite{cubic} for additional discussions on this issue. 

It is noted again that these two EL equations can be regarded as the field equations of the considered gravity,  which should be recovered exactly by the corresponding tensorial Einstein field equation. Apparently, they do follow the Bianchi identity as shown above. Therefore, Eqs. \eqref{first-field-equation} and \eqref{second-field-equation} can be interpreted as the $00$- and $ii$-components of the Einstein field equation of the sixth-order gravity.  
\subsection{Second-order limit}
In many cases, the second-order field equations are preferred. Perhaps, a main reason for this preference is due to the so-called Ostrogradsky ghost issue, whose appearance would lead to an instability. Phenomenologically, the late-time expanding phase of the universe  seems to be more suitable for the second-order field equations. As a result, by eliminating all higher-than-two-order terms in the above field equations, we can end up with the desired second-order field equations. To achieve such a thing, we need the help of constraints coming from the first EL equation \eqref{first-field-equation},
\begin{align}
\label{2nd-limit-1}
3\beta_1 +\beta_2 =&~0,\\
\label{2nd-limit-2}
3\gamma_1+\gamma_2 =&~0,\\
\label{2nd-limit-3}
-60\gamma_1 -16\gamma_2 +432\gamma_3 +120\gamma_4 +36\gamma_5 +96\gamma_7 +24\gamma_8 =&~0,\\
\label{2nd-limit-4}
-144\gamma_1 -36\gamma_2 +216\gamma_3 +72\gamma_4 +30\gamma_5 +72\gamma_7 +24\gamma_8 =&~0.
\end{align}
It is apparent that the first constraint equation has been long known within the  quadratic gravity, while the second one is understandable since $\gamma_1$ and $\gamma_2$ are coefficients of terms generating sixth-order derivatives. 

In addition, we also need the help of other constraints coming from the second EL equation \eqref{second-field-equation}, 
\begin{align}
\label{2nd-limit-5}
-36\gamma_1 +12 \gamma_2 +2592 \gamma_3 +720\gamma_4 +216\gamma_5 +576 \gamma_7 +144\gamma_8 =&~0,\\
\label{2nd-limit-6}
-996 \gamma_1 -244 \gamma_2 +3024 \gamma_3 +912 \gamma_4 +324 \gamma_5 +816 \gamma_7 +240 \gamma_8 =&~0,\\
\label{2nd-limit-7}
-162\gamma_1-42\gamma_2+216\gamma_3 +72\gamma_4 +30\gamma_5 +72\gamma_7 +24\gamma_8 =&~0,\\
\label{2nd-limit-8}
-276\gamma_1 -88\gamma_2 +432\gamma_3 +120\gamma_4 +36\gamma_5 +96\gamma_7 +24\gamma_8 =&~0,\\
\label{2nd-limit-9}
-252\gamma_1 -72\gamma_2 +216\gamma_3 +72\gamma_4 +30\gamma_5 +72\gamma_7 +24\gamma_8 =&~0.
\end{align}
As a result, Eqs. \eqref{2nd-limit-3} and \eqref{2nd-limit-8} both imply a solution,
\begin{equation}
\label{2nd-limit-10}
\gamma_1 =\gamma_2 =0.
\end{equation}
Interestingly, this solution can also be figured out from three other equations \eqref{2nd-limit-4}, \eqref{2nd-limit-7}, and \eqref{2nd-limit-9}. In addition, this solution does satisfy Eq. \eqref{2nd-limit-2}.

Mathematically,  Eqs. \eqref{2nd-limit-3}, \eqref{2nd-limit-4}, \eqref{2nd-limit-5}, \eqref{2nd-limit-6}, \eqref{2nd-limit-7}, \eqref{2nd-limit-8}, \eqref{2nd-limit-9} can be reduced, thanks to this solution, to a homogeneous set of  linear  equations given by
\begin{align}
\label{2nd-limit-11}
36 \gamma_3 +10 \gamma_4 +3 \gamma_5 + 8 \gamma_7 +2 \gamma_8 =&~0,\\
\label{2nd-limit-12}
36 \gamma_3 +12 \gamma_4 +5 \gamma_5 +12 \gamma_7 +4\gamma_8 =&~0.
\end{align}
As a result, the field equation \eqref{first-field-equation} will be reduced, in this limit, to the following form given by
\begin{align} \label{2nd-limit-13}
& \alpha \dot\beta^2   -36 \gamma_3  \left(\ddot\beta+2 {\dot\beta}^2\right) \left(2 {\ddot\beta}^2-19 {\dot\beta}^2 \ddot\beta+2 {\dot\beta}^4\right) -12\gamma_4 \left(2 {\ddot\beta}^3-14 {\dot\beta}^2 {\ddot\beta}^2 - 30 {\dot\beta}^4 \ddot\beta +3 {\dot\beta}^6 \right)\nonumber\\
&  - \gamma_5 \left( 10 {\ddot\beta}^3-63 {\dot\beta}^2 {\ddot\beta}^2 -108 {\dot\beta}^4 \ddot\beta +9 {\dot\beta}^6\right) -12\gamma_7 \left( 2 {\ddot\beta}^3-13 {\dot\beta}^2 {\ddot\beta}^2 -24 {\dot\beta}^4 \ddot\beta +2 {\dot\beta}^6 \right) \nonumber\\
& -4\gamma_8 \left( 2 {\ddot\beta}^3-12 {\dot\beta}^2 {\ddot\beta}^2  -18 {\dot\beta}^4 \ddot\beta + {\dot\beta}^6 \right) =0.
\end{align} 
Very interestingly, this equation can still be reduced a simpler form, in which all terms involving the second-order time derivative will disappear due to the constraint equations \eqref{2nd-limit-11} and \eqref{2nd-limit-12} as well as their suitable linear combination(s),
\begin{equation} \label{limit-2nd-limit-13}
 \alpha \dot\beta^2   - \left(144 \gamma_3+ 36\gamma_4 +9\gamma_5+ 24\gamma_7 +4\gamma_8 \right) \dot\beta^6  =0.
\end{equation}
On the other hand, the field equation \eqref{second-field-equation} will be reduced, in this limit, to the following form given by
\begin{align} \label{2nd-limit-14}
&\alpha \left(2 \ddot\beta+3 {\dot\beta}^2 \right) +108\gamma_3 \left( 8 {\ddot\beta}^3 +63 {\dot\beta}^2 {\ddot\beta}^2    +28 {\dot\beta}^4 \ddot\beta -4 {\dot\beta}^6\right) \nonumber\\
& +12\gamma_4 \left(22 {\ddot\beta}^3 +162 {\dot\beta}^2 {\ddot\beta}^2+72 {\dot\beta}^4 \ddot\beta  -9 {\dot\beta}^6  \right)  +3\gamma_5 \left(32 {\ddot\beta}^3+207 {\dot\beta}^2 {\ddot\beta}^2 +90 {\dot\beta}^4 \ddot\beta -9 {\dot\beta}^6\right) \nonumber\\
& +12 \gamma_7 \left(20 {\ddot\beta}^3+135 {\dot\beta}^2 {\ddot\beta}^2 +60 {\dot\beta}^4 \ddot\beta -6 {\dot\beta}^6  \right)  +12\gamma_8 \left(6 {\ddot\beta}^3+36 {\dot\beta}^2 {\ddot\beta}^2   +16 {\dot\beta}^4 \ddot\beta -{\dot\beta}^6\right)=0.
\end{align}
Similarly, this equation can be further reduced to
\begin{align} \label{limit-2nd-limit-14}
&\alpha \left(2 \ddot\beta+3 {\dot\beta}^2 \right) +432 \gamma_3 \left( 7 {\dot\beta}^4 \ddot\beta - {\dot\beta}^6\right)  +108 \gamma_4 \left(8 {\dot\beta}^4 \ddot\beta  - {\dot\beta}^6  \right) \nonumber\\
& +27 \gamma_5 \left(10 {\dot\beta}^4 \ddot\beta - {\dot\beta}^6\right)  +72 \gamma_7 \left(10 {\dot\beta}^4 \ddot\beta - {\dot\beta}^6  \right)  +12\gamma_8 \left(16 {\dot\beta}^4 \ddot\beta -{\dot\beta}^6\right)=0,
\end{align}
in which all terms containing $\ddot\beta^3$ or $\ddot\beta^2$ have been eliminated automatically, also due to the constraint equations \eqref{2nd-limit-11} and \eqref{2nd-limit-12} as well as their suitable linear combination(s).

In conclusion, the second-order limit requires the following constraints among field parameters, which have been pointed out in Eqs. \eqref{2nd-limit-1}, \eqref{2nd-limit-10}, \eqref{2nd-limit-11}, and \eqref{2nd-limit-12}. Mathematically, these constraint equations form a homogeneous set of  linear  equations, which admits infinitely many non-trivial solutions of $\beta_i~(i=1-2)$ and $\gamma_j~(j=3, ~4,~5,~7,~8)$. Two field equations in this limits are given by Eqs. \eqref{limit-2nd-limit-13} and \eqref{limit-2nd-limit-14}, in which one of them is the second-order ODE. Importantly, they do follow the Bianchi identity as expected.
\subsection{Fourth-order limit}
In the inflationary sense, fourth-order gravities seem to be more favorable \cite{Schmidt:2006jt,Alvarez-Gaume:2015rwa,Salvio:2018crh}, despite the fact that they could contain the Ostrogradsky ghost \cite{Woodard:2015zca}. A typical and motivating example is the well-known Starobinsky model \cite{Starobinsky:1980te}. It should be noted that the mass of the Ostrogradsky ghost in ghostful fourth-order gravity models/theories should be sufficiently heavier than the inflation scale in order to ensure the predictability of these models/theories \cite{Aoki:2019snr,Lambiase:2025qyl}.

Generally, fourth-order gravities are expected to admit stable quasi-de Sitter solutions \cite{Elizalde:2014xva,Pozdeeva:2019agu,Vernov:2021hxo} and therefore do not face the so-called eternal inflation issue.  This means that exact de Sitter solutions should not exist in fourth-order gravities or if exist they should be unstable  during the inflationary phase. To achieve the fourth-order limit, we must turn off all the fifth- and sixth-order time derivatives in the field equations (a.k.a. the EL equations). This requirement addresses  the following simple constraint,
\begin{equation} \label{4-order-constraint}
3\gamma_1 +\gamma_2 =0,
\end{equation}
by which we will have the corresponding field equations given by
\begin{align}  \label{4th-limit-1}
& \alpha \dot\beta^2 + 2\left( 3 \beta_1 +\beta_2 \right)  \left(2\dot\beta  \beta ^{(3)}  -{\ddot\beta}^2+6 {\dot\beta}^2 \ddot\beta \right) \nonumber\\
& -6 \gamma_1  \left(24\dot\beta \ddot\beta \beta ^{(3)}+10{\dot\beta}^3 \beta ^{(3)}  -8 {\ddot\beta}^3+32 {\dot\beta}^2 {\ddot\beta}^2 -24 {\dot\beta}^4 \ddot\beta \right) \nonumber\\
& - 2\gamma_2  \left( 18\dot\beta \ddot\beta \beta ^{(3)}+8 {\dot\beta}^3 \beta ^{(3)}  - 6 {\ddot\beta}^3 +22 {\dot\beta}^2 {\ddot\beta}^2-30 {\dot\beta}^4 \ddot\beta \right) \nonumber\\
& +36 \gamma_3  \left(\ddot\beta+2 {\dot\beta}^2\right) \left(6  \dot\beta \beta ^{(3)} -2 {\ddot\beta}^2+19 {\dot\beta}^2 \ddot\beta-2 {\dot\beta}^4\right) \nonumber\\
& +12\gamma_4 \left(6  \dot\beta \ddot\beta \beta ^{(3)} +10{\dot\beta}^3 \beta ^{(3)}-2 {\ddot\beta}^3+14 {\dot\beta}^2 {\ddot\beta}^2 +30 {\dot\beta}^4 \ddot\beta -3 {\dot\beta}^6 \right) \nonumber\\
& +\gamma_5 \left(30\dot\beta \ddot\beta \beta ^{(3)} +36{\dot\beta}^3 \beta ^{(3)}  -10 {\ddot\beta}^3+63 {\dot\beta}^2 {\ddot\beta}^2 +108 {\dot\beta}^4 \ddot\beta -9 {\dot\beta}^6\right) \nonumber\\
&+12\gamma_7 \left(6\dot\beta \ddot\beta \beta ^{(3)} +8 {\dot\beta}^3 \beta ^{(3)} -2 {\ddot\beta}^3+13 {\dot\beta}^2 {\ddot\beta}^2 +24 {\dot\beta}^4 \ddot\beta -2 {\dot\beta}^6 \right) \nonumber\\
& +4\gamma_8 \left( 6\dot\beta \ddot\beta \beta ^{(3)}+ 6 {\dot\beta}^3 \beta ^{(3)} -2 {\ddot\beta}^3+12 {\dot\beta}^2 {\ddot\beta}^2  +18 {\dot\beta}^4 \ddot\beta - {\dot\beta}^6 \right) =0
\end{align}
and
\begin{align}  \label{4th-limit-2}
&\alpha \left(2 \ddot\beta+3 {\dot\beta}^2 \right) + 2\left(3 \beta_1 +\beta_2 \right)  \left(2 \beta ^{(4)}+12\dot\beta \beta ^{(3)} +9 {\ddot\beta}^2+18 {\dot\beta}^2 \ddot\beta\right) \nonumber\\
& - 6\gamma_1 \left(42\ddot\beta \beta ^{(4)} +46{\dot\beta}^2  \beta ^{(4)}  +27 {\beta ^{(3)}}^2  +166 \dot\beta \ddot\beta \beta ^{(3)}  +6{\dot\beta}^3 \beta ^{(3)} +40 {\ddot\beta}^3 -72 {\dot\beta}^4 \ddot\beta \right) \nonumber\\
& -2\gamma_2  \left(36 \ddot\beta \beta ^{(4)} +44 {\dot\beta}^2 \beta ^{(4)}+21 {\beta ^{(3)}}^2 +122\dot\beta \ddot\beta \beta ^{(3)}   -6 {\dot\beta}^3 \beta ^{(3)}  +26 {\ddot\beta}^3 -54 {\dot\beta}^2 {\ddot\beta}^2 -90 {\dot\beta}^4 \ddot\beta \right) \nonumber\\
& +108\gamma_3 \left( 2\ddot\beta  \beta ^{(4)} +4{\dot\beta}^2 \beta ^{(4)} +2 {\beta ^{(3)}}^2+28\dot\beta \ddot\beta  \beta ^{(3)} +24{\dot\beta}^3 \beta ^{(3)}  +8 {\ddot\beta}^3 +63 {\dot\beta}^2 {\ddot\beta}^2    +28 {\dot\beta}^4 \ddot\beta -4 {\dot\beta}^6\right) \nonumber\\
& +12\gamma_4 \left(6\ddot\beta  \beta ^{(4)} +10{\dot\beta}^2 \beta ^{(4)}+ 6 {\beta ^{(3)}}^2 +76\dot\beta \ddot\beta \beta ^{(3)} +60 {\dot\beta}^3  \beta ^{(3)}  +22 {\ddot\beta}^3 +162 {\dot\beta}^2 {\ddot\beta}^2+72 {\dot\beta}^4 \ddot\beta  -9 {\dot\beta}^6  \right) \nonumber\\
& +3\gamma_5 \left(10 \ddot\beta  \beta ^{(4)}+12{\dot\beta}^2 \beta ^{(4)}+10 {\beta ^{(3)}}^2+108\dot\beta \ddot\beta \beta ^{(3)}  +72{\dot\beta}^3 \beta ^{(3)} +32 {\ddot\beta}^3+207 {\dot\beta}^2 {\ddot\beta}^2 +90 {\dot\beta}^4 \ddot\beta -9 {\dot\beta}^6\right) \nonumber\\
& +12 \gamma_7 \left(6\ddot\beta \beta ^{(4)} +8 {\dot\beta}^2 \beta ^{(4)} +6 {\beta ^{(3)}}^2+68\dot\beta \ddot\beta \beta ^{(3)}+48{\dot\beta}^3 \beta ^{(3)}  +20 {\ddot\beta}^3+135 {\dot\beta}^2 {\ddot\beta}^2 +60 {\dot\beta}^4 \ddot\beta -6 {\dot\beta}^6  \right) \nonumber\\
& +12\gamma_8 \left(2\ddot\beta \beta ^{(4)} +2{\dot\beta}^2 \beta ^{(4)}+2 {\beta ^{(3)}}^2+20\dot\beta \ddot\beta \beta ^{(3)} +12{\dot\beta}^3 \beta ^{(3)}   +6 {\ddot\beta}^3+36 {\dot\beta}^2 {\ddot\beta}^2   +16 {\dot\beta}^4 \ddot\beta -{\dot\beta}^6\right)=0.
\end{align}
It is noted that these two field equations cannot be further reduced to simpler forms since we only have one constraint equation \eqref{4-order-constraint}. 
It is apparent that only $\gamma_1$ and $\gamma_2$ are tightly constrained by Eq. \eqref{4-order-constraint}, while the other parameters like $\beta_i~(i=1-2)$ and $\gamma_j~(j=3, ~4,~5,~7,~8)$ are unconstrained. One can easily check that these two field equations do obey the Bianchi identity as expected. And one of them is clearly the fourth-order ODE.
 \section{Stability investigation of the de Sitter solution} \label{sec3}
 \subsection{de Sitter solution}
In this subsection, we would like to find out de Sitter solutions to the sixth-order gravity with $3\gamma_1 +\gamma_2 \neq 0$  by using the following ansatz, 
\begin{equation}
\beta(t) =\zeta t,
\end{equation}
where $\zeta$ is an undetermined parameter, which will be determined from the field equations. As a result, it simply turns out for this ansatz that 
\begin{equation}
\dot\beta =\zeta, \quad \ddot\beta =\beta^{(3)}=\beta^{(4)}=\beta^{(5)}=\beta^{(6)} =0.
\end{equation}
Consequently, both field equations derived above, i.e., Eqs. \eqref{first-field-equation} and \eqref{second-field-equation}, will reduce to the same simple algebraic equation of $\zeta$,
\begin{equation} \label{equation-of-zeta}
\left(144\gamma_3 +36\gamma_4 +9\gamma_5 +24\gamma_7 +4\gamma_8  \right)\zeta^4 -\alpha =0.
\end{equation}
Hence, a simple solution of this equation can be solved to be
\begin{equation}\label{value-of-zeta}
\zeta = \left(\frac{\alpha}{144\gamma_3 +36\gamma_4 +9\gamma_5 +24\gamma_7 +4\gamma_8} \right)^{1/4}.
\end{equation}
Since $\alpha$ is always positive definite in the Einstein's gravity (it is normally set to be one in many scenarios), the real positive value of $\zeta$, which corresponds to a de Sitter universe, will put a constraint on the other parameters such as
\begin{equation} \label{inequality-1}
144\gamma_3 +36\gamma_4 +9\gamma_5 +24\gamma_7 +4\gamma_8  >0.
\end{equation}

Once again, it is interesting to note that the well-known $R^2$ term, which is populated thanks to the Starobinsky inflationary model \cite{Starobinsky:1980te}, does not contribute anything to the value of the obtained de Sitter solution. Similar argument is applied to $\beta_2$-, $\gamma_1$-, and $\gamma_2$-terms. This result indicates that the terms generating the sixth-order time derivative in the field equations do not prefer the existence of de Sitter solution, similar to the quadratic curvature terms $R^2$ and $R_{\mu\nu}^2$. As will be shown later, however, they will actively affect on the stability of the obtained de Sitter solution. 
 \subsection{Dynamical system}
 As said above, we would like to address in this section one of the most important issues of cosmological solutions, which is nothing but their stability. To do this task, we will use the powerful method based on the dynamical system \cite{Barrow:2006xb,Do:2023yvg,Pham:2024fub,Bahamonde:2017ize}. In particular, we will follow the approach used in our recent paper on the generalized Einsteinian cubic gravity \cite{cubic}, in which Eq. \eqref{second-field-equation} will be used, while Eq. \eqref{first-field-equation} acts as a constraint one. Inversely, as demonstrated in our paper \cite{cubic}, Eq. \eqref{first-field-equation} can be used for the stability analysis, while  Eq. \eqref{second-field-equation}  remains as a constraint one. A reason for this additional approach is due to the fact that Eq.  \eqref{second-field-equation} is the differential consequence of Eq. \eqref{first-field-equation} as shown above. An interesting example can be seen in Ref. \cite{Toporensky:2006kc}.  Before going to derive the corresponding autonomous equations for the sixth-order gravity, it is important to remark that there is a simpler stability analysis method, which is not based on the dynamical system, e.g., see Ref. \cite{CamposDelgado:2024jst} for a recent related work. In particular, this method requires a direct perturbation of the scale factor $\beta(t)$ around the de Sitter solution and  concerns only Eq. \eqref{first-field-equation}. To be complete and to be consistent with Ref. \cite{cubic}, we will consider this method as an important cross-check in Appendix \ref{app} in order to ensure that the stability analysis performed in this section leads to correct results.   
 
 First, we need to introduce dynamical variables for the sixth-order gravity as follows
 \begin{equation}
B=\frac{1}{\dot\beta^2},\quad Q=\frac{\ddot\beta}{\dot\beta^2},\quad Q_2 =\frac{\beta^{(3)}}{\dot\beta^3}, \quad Q_3 =\frac{\beta^{(4)}}{\dot\beta^4}, \quad Q_4 =\frac{\beta^{(5)}}{\dot\beta^5}.
 \end{equation}
 Consequently, their autonomous equations read
 \begin{align}
  \label{Dyn-1}
 B' &= -2QB,\\
  \label{Dyn-2}
 Q' &=Q_2 -2Q^2,\\
 \label{Dyn-3}
 Q_2 ' & =  Q_3 -3Q Q_2,\\
 \label{Dyn-4}
 Q_3' & = Q_4 -4 Q Q_3 , \\
 \label{Dyn-5}
 Q_4 ' &= \frac{\beta^{(6)}}{\dot\beta^6} - 5 Q Q_4.
 \end{align}
Here, the prime is understood as a derivative w.r.t. the dynamical time variable  $\tau \equiv \int \dot\beta dt$, e.g., $B' \equiv dB/d\tau$. This set of autonomous equations will be fully defined if the remaining term in the last equation, i.e., $\frac{\beta^{(6)}}{\dot\beta^6}$, is determined. It turns out that this term will be solely defined from  field equations of sixth-order gravity. Different setups of sixth-order gravity will lead to different field equations and therefore different expressions of $\frac{\beta^{(6)}}{\dot\beta^6}$ and  dynamical systems.

Rewriting Eq. \eqref{second-field-equation} in terms of the introduced dynamical variables will help us to figure out the desired expression of $\frac{\beta^{(6)}}{\dot\beta^6}$  to be
\begin{align} \label{Dyn-6}
\frac{\beta^{(6)}}{\dot\beta^6} =& ~ \frac{1}{4\left( 3\gamma_1 +\gamma_2 \right)} \nonumber\\
&\times  \left[ \alpha B^2 \left( 2 Q +3 \right) \right. \nonumber\\
&\left. +2 \left(3\beta_1+\beta_2 \right) B \left( 2 Q_3 +12 Q_2  + 9 Q^2  +18 Q \right)  \right. \nonumber\\
&\left. -6\gamma_1 \left(18 Q_4+ 42 Q Q_3 +46 Q_3 +27 Q_2^2+166 Q Q_2  + 6Q_2 +40 Q^3 -72 Q   \right) \right. \nonumber\\
&\left. -2 \gamma_2 \left(18 Q_4+36Q Q_3 +44 Q_3 +21 Q_2^2 +122 Q Q_2  -6 Q_2 +26 Q^3  -54 Q^2 -90 Q    \right) \right. \nonumber\\
&\left. +108\gamma_3 \left(2Q Q_3  +4Q_3 + 2 Q_2^2 +28 Q Q_2 +24 Q_2 +8 Q ^3  +63 Q^2  +28 Q -4    \right) \right. \nonumber\\ 
&\left. +12\gamma_4 \left( 6Q Q_3 +10 Q_3  + 6 Q_2^2+ 76 Q Q_2  +60 Q_2 + 22 Q^3 +162 Q^2 +72 Q    - 9   \right) \right. \nonumber\\
&\left. + 3\gamma_5 \left(10 Q Q_3 + 12 Q_3 + 10 Q_2^2 + 108 Q Q_2  +72 Q_2 +32 Q^3 + 207 Q^2+ 90Q  - 9   \right) \right. \nonumber\\
& \left. + 12\gamma_7 \left(6Q Q_3 +8 Q_3+6Q_2^2 + 68 Q Q_2 +48 Q_2 + 20 Q^3+135 Q^2 +60 Q   -6    \right) \right. \nonumber\\
&\left. +12\gamma_8 \left(2 Q Q_3 +2 Q_3+2 Q_2^2 +20 Q Q_2+12 Q_2  +6Q^3 +36 Q^2  +16 Q  -1  \right) \right].
\end{align}
It is important to note that there exists a constraint equation coming from the Friedmann equation \eqref{first-field-equation},
\begin{align} \label{constraint-eq}
& \alpha B^2 +2\left(3\beta_1 +\beta_2 \right) B \left(2 Q_2 -  Q^2 +6  Q \right) \nonumber\\
&-6\gamma_1 \left(2Q_4 -2QQ_3 +12 Q_3 +Q_2^2 +24 QQ_2 +10 Q_2 -8Q^3+32 Q^2 -24 Q \right) \nonumber\\
&-2\gamma_2 \left(2Q_4 -2QQ_3 +12 Q_3 +Q_2^2 +18 QQ_2 +8 Q_2 -6Q^3+22 Q^2 -30 Q \right) \nonumber\\
&+36\gamma_3 \left( Q +2\right)\left(6Q_2 -2Q^2 +19 Q -2\right) \nonumber\\
&+12\gamma_4 \left(6 Q Q_2 +10Q_2 -2Q^3 +14 Q^2 +30 Q -3  \right) \nonumber\\
&+\gamma_5 \left(30 Q Q_2 +36Q_2 -10Q^3 +63 Q^2 +108 Q -9  \right) \nonumber\\
&+12\gamma_7 \left(6 Q Q_2 +8Q_2 -2Q^3 +13 Q^2 +24 Q -2  \right) \nonumber\\
&+4\gamma_8 \left(6 Q Q_2 +6Q_2 -2Q^3 +12 Q^2 +18 Q -1  \right) =0.
\end{align}
Before ending this subsection, it is worth noting that one can use this constraint Friedmann equation \eqref{constraint-eq}  rather than Eq. \eqref{Dyn-6} for investigating fixed points along with their stability, similar to what we have demonstrated  in our very recent paper \cite{cubic}. A reason for this is due to the fact that this equation comes from the $00$-component of the Einstein field equation shown in Eq. \eqref{first-field-equation}, whose differential consequence is nothing but the $ii$-component of the Einstein field equation, i.e., Eq. \eqref{second-field-equation}. 
\subsection{de Sitter fixed point}
Now, we are going to seek a de Sitter fixed point of the dynamical system, which is expected to be equivalent to the de Sitter solution found above. Mathematically, fixed points of the dynamical system are solutions of the following set of equations,
\begin{equation}
B' =Q'=Q_2'=Q_3' =Q_4' =0 .
\end{equation}
According to Eq. \eqref{Dyn-1}, $B'=0$ will imply two possible solutions, $Q=0$ or $B=0$. It appears that the last solution, i.e., $B=0$, does not correspond to the de Sitter solution with $B\neq 0$, which we have been interested in. Therefore, we will ignore it and prefer another solution, i.e., $Q=0$, which is consistent with the de Sitter solution for further stability analysis. As a consequence, we have the following solution, 
\begin{equation}
 \frac{\beta^{(6)}}{\dot\beta^6} =Q_4=Q_3 =Q_2 = 0,
\end{equation}
which is again consistent with the de Sitter solution. As a result, the solution $\frac{\beta^{(6)}}{\dot\beta^6} =0$ will lead Eq. \eqref{Dyn-6} to the corresponding equation of $B$,
\begin{equation} \label{equation-of-B}
\alpha B^2 -144 \gamma_3 - 36 \gamma_4 -9 \gamma_5 -24 \gamma_7 -4\gamma_8 =0,
\end{equation}
which is consistent with the Eq. \eqref{equation-of-zeta} and can be recovered from the constraint equation \eqref{constraint-eq}. For a non-vanishing $\alpha$, it turns out that
\begin{equation}
B^2 = \frac{1}{\alpha}\left( 144 \gamma_3 + 36 \gamma_4 + 9 \gamma_5 + 24 \gamma_7 +4\gamma_8\right).
\end{equation}
Indeed, one can easily figure out from Eq. \eqref{equation-of-B}  that
\begin{equation}
\beta(t) =\zeta t ,
\end{equation}
with $\zeta$ being defined in Eq. \eqref{value-of-zeta}. This result clearly confirms our expectation that the considered fixed point with $B\neq 0$ is indeed equivalent to  the de Sitter solution derived in the previous subsection. For convenience, we will call it the de Sitter fixed point from now on.
\subsection{Non-de Sitter fixed points}
We would like to take a moment to discuss briefly about non-de Sitter fixed points corresponding to $B=0$, or equivalently $\dot\beta^2 = \infty$. Due to this result, one can call them unphysical fixed points.  As a consequence, we have the corresponding relations for the non-de Sitter fixed points,
\begin{equation}
Q_2 =2Q^2, \quad Q_3 = 6 Q^3, \quad Q_4 = 24 Q^4, \quad \frac{\beta^{(6)}}{\dot\beta^6} = 120 Q^5.
\end{equation}
Furthermore, either Eq. \eqref{Dyn-6} or Eq. \eqref{constraint-eq} can now be reduced to
\begin{equation} \label{non-de-Sitter-fixed-point}
\left(2Q+1 \right) \left(\kappa_4 Q^4+\kappa_3 Q^3+\kappa_2 Q^2+\kappa_1 Q+\kappa_0 \right)=0,
\end{equation}
where the coefficients $\kappa_i$ $(i=0-4)$ are given by
\begin{align}
\kappa_4 =& ~80 \left(3\gamma_1+\gamma_2  \right),\\
\kappa_3=& ~ 2 \left(336 \gamma _1+102 \gamma _2-180 \gamma _3-60 \gamma _4-25 \gamma _5-60 \gamma _7-20 \gamma _8\right),\\
\kappa_2=&~ 312 \gamma _1+76 \gamma _2-1404 \gamma _3-408 \gamma _4-135 \gamma _5-348 \gamma _7-96 \gamma _8,\\
\kappa_1 =&-12 \left(12 \gamma _1+5 \gamma _2+108 \gamma _3+30 \gamma _4+9 \gamma _5+24 \gamma _7+6 \gamma _8\right),\\
\kappa_0 =&~144 \gamma _3+36 \gamma _4+9 \gamma _5+24 \gamma _7+4 \gamma _8.
\end{align}
In principle, this equation always admits $Q=-1/2$ as its solution, which is independent of the parameters $\gamma_i~ (i=1, ~2,~3, ~4,~5,~7,~8)$. On the other hand, it also admits four other solutions, whose values will depend on the values of $\gamma_i$.
\subsection{Stability of the de Sitter fixed point} \label{stability}
Since we have only been interested in the de Sitter solution, we will focus on investigating its stability to see if it is compatible with the inflationary phase of  early universe. Normally, we will perturb the dynamical system around the de Sitter fixed point and see how perturbation modes involve with the dynamical time $\tau$ \cite{Bahamonde:2017ize}. It is known that a fixed point is stable agains perturbations if these perturbations all  tend to vanish as the dynamical time goes to infinity. Conversely, if at least one of the perturbations blows up as the dynamical time becomes large, then the corresponding fixed point is unstable.

Now, we would like to perturb the dynamical system by taking perturbations of dynamical variables such as 
\begin{equation}
B \to B+\delta B, \quad Q \to Q+\delta Q, \quad Q_2 \to Q_2 +\delta Q_2,\quad Q_3 \to Q_3+\delta Q_3, \quad Q_4 \to Q_4+\delta Q_4.
\end{equation}
As a consequence, the autonomous equations will be perturbed around the de Sitter fixed point as follows
\begin{align}
  \label{pert-1}
 \delta B' &= -2 B \delta Q,\\
  \label{pert-2}
\delta Q' &= \delta Q_2 ,\\
 \label{pert-3}
 \delta Q_2 ' & =  \delta Q_3 ,\\
 \label{pert-4}
 \delta Q_3' & = \delta Q_4  , \\
 \label{pert-5}
\delta  Q_4 ' &= \delta \left( \frac{\beta^{(6)}}{\dot\beta^6}\right),
 \end{align}
 where $\delta \left( \frac{\beta^{(6)}}{\dot\beta^6}\right)$ will be defined from a perturbed version of Eq. \eqref{Dyn-6}. In particular, it turns out that
 \begin{align} \label{perturbation-medium}
 \delta \left( \frac{\beta^{(6)}}{\dot\beta^6}\right) =&~ \frac{1}{4\left( 3\gamma_1 +\gamma_2 \right)} \nonumber\\
 &\times \left[ 6\alpha B \delta B +2\alpha B^2 \delta Q +2\left(3\beta_1+\beta_2 \right) B \left( 2\delta Q_3 + 12 \delta Q_2 +18\delta Q\right) \right. \nonumber\\
 & \left. -6\gamma_1 \left(18\delta Q_4 + 46\delta Q_3 -6\delta Q_2 -72\delta Q \right)  -2\gamma_2 \left(18\delta Q_4 +44\delta Q_3 -6\delta Q_2 -90\delta Q \right) \right. \nonumber\\
 & \left. +108\gamma_3 \left(4\delta Q_3 +24\delta Q_2 +28\delta Q \right) +12\gamma_4 \left(10\delta Q_3 +60\delta Q_2 +72\delta Q \right) \right. \nonumber\\
 &\left.+ 3\gamma_5 \left(12\delta Q_3 +72\delta Q_2 +90\delta Q \right) +12\gamma_7 \left(8\delta Q_3 +48\delta Q_2 +60\delta Q \right) \right. \nonumber\\
 &\left. +12\gamma_8 \left(2\delta Q_3 +12\delta Q_2+16\delta Q \right)\right].
 \end{align}
Fortunately,  this expression can be further simplified thanks to the perturbed version of the constraint equation \eqref{constraint-eq}, which can be defined to be
 \begin{align}
&  2\alpha B \delta B -4 \left(3\gamma_1+\gamma_2 \right) \left(\delta Q_4 +6 \delta Q_3  -25 \delta Q \right) \nonumber\\
 &+4 \left[ \left(3\beta_1 +\beta_2 \right)B - 15\gamma_1 - 4\gamma_2  +108 \gamma_3 +30\gamma_4 +9 \gamma_5+24 \gamma_7+6\gamma_8  \right] \left( \delta Q_2 +3\delta Q \right) =0.
 \end{align}
 Indeed, taking out $\alpha B \delta B$ from this equation and inserting the obtained result into Eq. \eqref{perturbation-medium} gives
 \begin{align}
  \delta \left( \frac{\beta^{(6)}}{\dot\beta^6}\right) = \frac{1}{4\left( 3\gamma_1 +\gamma_2 \right)} \left(q_3 \delta Q_3 +q_2 \delta Q_2 + q_1 \delta Q \right) -9\delta Q_4,
 \end{align}
 where
 \begin{align}
 q_3 & =  3 \beta _1 B+\beta _2 B-15 \gamma _1-4 \gamma _2+108 \gamma _3+30 \gamma _4+9 \gamma _5+24 \gamma _7+6 \gamma _8,\\
 q_2 &= 9 \beta _1 B+3 \beta _2 B+54 \gamma _1+15 \gamma _2+324 \gamma _3+90 \gamma _4+27 \gamma _5+72 \gamma _7+18 \gamma _8,\\
 q_1&=18 \gamma _1+6 \gamma _2-144 \gamma _3-36 \gamma _4-9 \gamma _5-24 \gamma _7-4 \gamma _8.
 \end{align}
 
  As usual, we take exponential perturbations,
 \begin{align}
 \delta B &= A_B e^{\lambda \tau}, \\
  \delta Q& = A_Q e^{\lambda \tau},\\
  \delta Q_2 &= A_{Q_2} e^{\lambda \tau}, \\
  \delta Q_3 &= A_{Q_3} e^{\lambda \tau}, \\
 \delta Q_4 &= A_{Q_4} e^{\lambda \tau}, 
 \end{align}
 where $\lambda$ is a parameter characterizing the stability of the de Sitter fixed point. In particular, if $\lambda$ is positive definite then all perturbations of dynamical variables will blow up as the dynamical time $\tau$ becomes large. Then, the corresponding de Sitter fixed point will be unstable. On the other hand, the de Sitter fixed point will be stable if $\lambda$ is negative definite since all perturbations of dynamical variables will approach zero as the dynamical time $\tau$ goes to infinity. Our goal now is to examine whether $\lambda$ is positive or not within the sixth-order gravity. 
 
 As a result, all perturbed equations shown above will reduce to the corresponding algebraic equations given by
  \begin{align}
 \lambda A_B =& -2B A_Q,\\
  \lambda  A_Q = &~A_{Q_2},\\
   \lambda  A_{Q_2} =&~ A_{Q_3},\\
    \lambda  A_{Q_3} =&~ A_{Q_4},\\
    \lambda  A_{Q_4} = &~ \frac{1}{4\left( 3\gamma_1 +\gamma_2 \right)} \left(q_3 A_{ Q_3} +q_2 A_{ Q_2} + q_1 A_ Q \right) -9A_{Q_4},
 \end{align}
 respectively. Mathematically, this set of homogeneous equations can be written in a matrix form such as
\begin{equation} \label{stability-equation}
 {\cal A}\left( {\begin{array}{*{20}c}
   A_B  \\
   A_Q  \\
   A_{Q_2} \\
   A_{Q_3}\\
   A_{Q_4}\\
 \end{array} } \right) \equiv \left[ {\begin{array}{*{20}c}
   {\lambda} & {2B} & {0 } &{0}&{0}  \\
   { 0} & {\lambda} & {-1 } &{0}&{0}  \\
     {0  } & { 0} & {  \lambda }&{-1}&{0}  \\
         {0  } & { 0} & {  0 }&{\lambda}&{-1}  \\
              {0  } & { -\frac{q_1}{4\left(3\gamma_1+\gamma_2 \right)}} & {  -\frac{q_2}{4\left(3\gamma_1+\gamma_2 \right)} }&{-\frac{q_3}{4\left(3\gamma_1+\gamma_2 \right)}}&{\lambda + 9}  \\
 \end{array} } \right]  \left( {\begin{array}{*{20}c}
   A_B  \\
   A_Q  \\
   A_{Q_2} \\
   A_{Q_3}\\
   A_{Q_4}\\
 \end{array} } \right) = 0,
\end{equation}
where $q_1$, $q_2$, and $q_3$ have been defined above.
Since we are looking for non-vanishing solutions of this set equation, the determinant of matrix ${\cal A}$ must be equal to zero, i.e, $\det {\cal A} =0$. As a result, this condition will lead to the corresponding equation of $\lambda$ defined as
\begin{equation} \label{polynomial-equation}
\lambda \left(a_4 \lambda^4 + a_3 \lambda^3 +a_3 \lambda^3 +a_2 \lambda^2 +a_1 \lambda +a_0 \right)=0,
\end{equation} 
where 
\begin{align}
a_4& = 4 \left(3 \gamma _1+\gamma _2\right),\\
a_3 &=9a_4 ,\\
a_2&= -\left(3 \beta _1+\beta_2 \right) B+15 \gamma _1+4 \gamma _2-108 \gamma _3-30 \gamma _4-9 \gamma _5-24 \gamma _7-6 \gamma _8,\\
a_1&=  -\frac{3}{4}a_4 -6 \left(15\gamma_1 +4\gamma_2 \right)+3a_2 ,\\
a_0&=-\frac{3}{2} a_4 +144 \gamma _3+36 \gamma _4+9 \gamma _5+24 \gamma _7+4 \gamma _8.
\end{align}
Interestingly, it is apparent that  $a_4$ will be negative definite  if 
\begin{equation} \label{inequality-2}
3\gamma_1+\gamma_2 <0.
\end{equation}
Consequently, $a_0$ will be positive definite since $144 \gamma _3+36 \gamma _4+9 \gamma _5+24 \gamma _7+4 \gamma _8>0$ as required to get the real de Sitter solution (or equivalently the real de Sitter fixed point), according to Eqs. \eqref{value-of-zeta} and \eqref{equation-of-B}.  Mathematically, it turns out  that Eq. \eqref{polynomial-equation} will therefore admit at least one positive root of $\lambda$ since $a_4 a_0 <0$. This important result clearly indicates that the corresponding de Sitter fixed point will become unstable. And in the light of multiverse associated with eternal inflation \cite{Guth:2007ng,Brustein:1994kw}, the sixth-order gravity admitting an unstable de Sitter solution would be relevant  to the inflationary phase of early universe.  

On the other hand, for $3\gamma_1+\gamma_2 >0$, or equivalently $a_4>0$ as well as $a_3>0$, then it is possible to have a stable de Sitter fixed point if all other coefficients $a_i ~(i=0-2)$ turn out to be positive. This is based on an observation that a polynomial equation will only have non-positive roots if its coefficients are all positive or negative definite. It is noted that a gravity model admitting a stable de Sitter solution seems to be more suitable for the late-time accelerated expansion of  universe. 
\subsection{Attractor property of the de Sitter fixed point}
\begin{figure}[hbtp] 
 \centering
	  \includegraphics[scale=0.5]{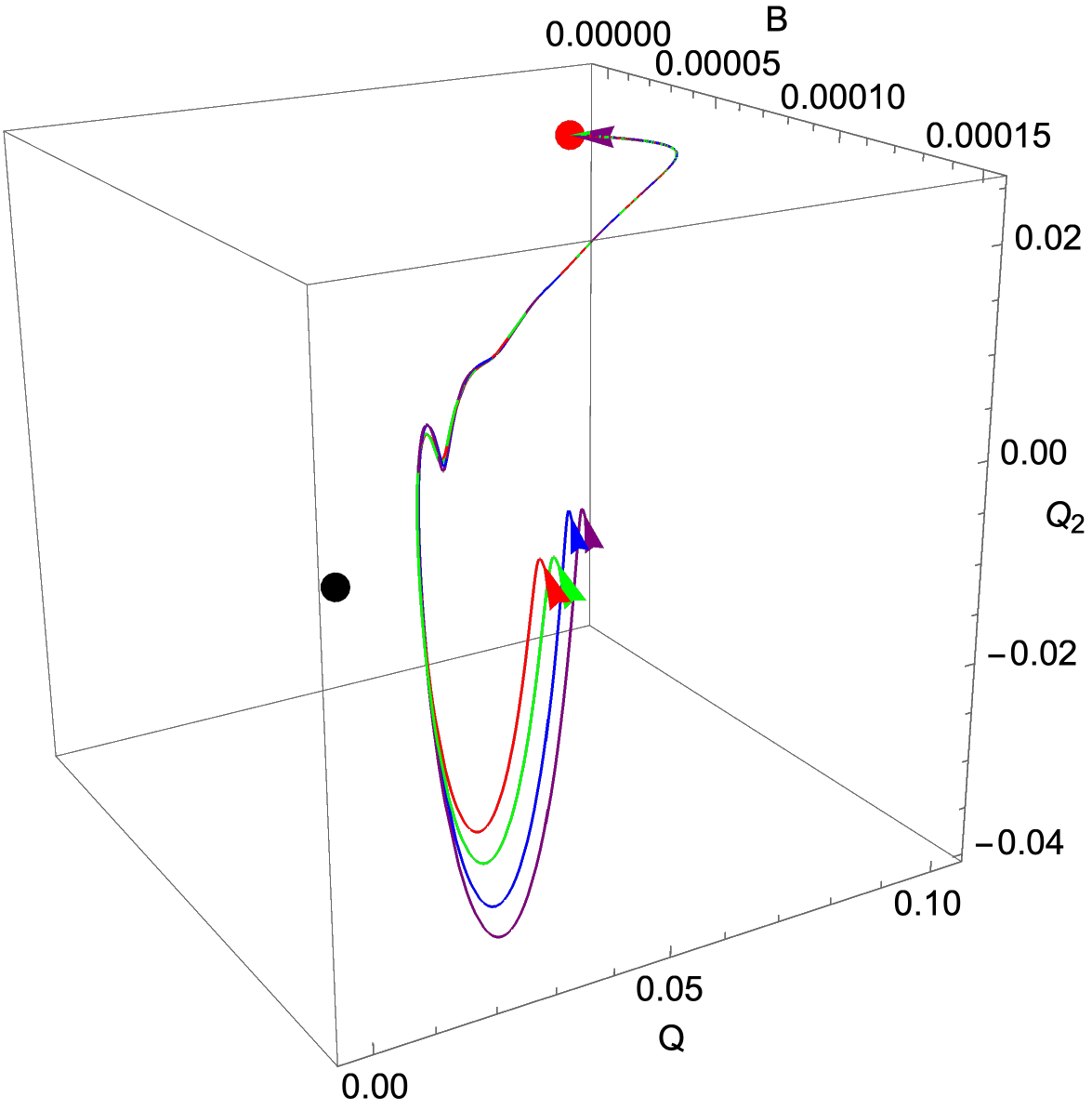}\\
\caption{\it The de Sitter fixed point with $\left(B,Q,Q_2\right)\simeq \left(1.473\times 10^{-4},0,0\right)$, displayed as a black point, acts as an repeller. An attractor with $\left(B,Q,Q_2\right) \simeq \left(0,9.826 \times 10^{-2}, 1.931 \times 10^{-2} \right)$, displayed as a red point, corresponds to a non-de Sitter fixed point associated with a real solution $Q$ of Eq. \eqref{non-de-Sitter-fixed-point}. }
\label{fig1}
\end{figure}

In harmony with the stability analysis done in the previous subsection, we would like to see if the de Sitter fixed point is an attractor or not. To do this, we are going to numerically solve the dynamical system for different initial conditions to see an evolution of the corresponding trajectories in a phase space of $B$, $Q$, and $Q_2$. In the dynamical system context, an attractor is a fixed point that different trajectories with different initial conditions all tend to converge to. In contrast, a repeller is a fixed point that all trajectories tend to repel. We will set  $\alpha=1$, $\beta_1=\beta_2 =-10^{-5}$, $\gamma_1 =\gamma_2 =-10^{-10}$, and $\gamma_3=\gamma_4=\gamma_5=\gamma_7=\gamma_8=10^{-10}$, for which all inequalities shown in Eqs. \eqref{inequality-1} and \eqref{inequality-2} are satisfied. Then, we numerically solve the corresponding dynamical system with four different initial conditions and plot the corresponding trajectories in the phase space for a detailed visualization. In particular, the red, green, blue, and purple trajectories displayed in Fig. \ref{fig1} respectively correspond to initial conditions such as\\
$\bullet$ red curve: $\left(B[0],Q[0],Q_2[0],Q_3[0],Q_4[0] \right)= \left(1.475 \times 10^{-4}, 3.2 \times 10^{-2}, -2\times 10^{-3}, 10^{-2}, 10^{-2}\right)$,\\
$\bullet$ green curve: $\left(B[0],Q[0],Q_2[0],Q_3[0],Q_4[0] \right)= \left(1.48 \times 10^{-4}, 3.4 \times 10^{-2}, -2\times 10^{-3}, 10^{-2}, 10^{-2}\right)$,\\
$\bullet$ blue curve: $\left(B[0],Q[0],Q_2[0],Q_3[0],Q_4[0] \right)= \left(1.485 \times 10^{-4}, 3.6 \times 10^{-2}, 2\times 10^{-3}, 10^{-2}, 10^{-2}\right)$,\\
$\bullet$ purple curve: $\left(B[0],Q[0],Q_2[0],Q_3[0],Q_4[0] \right)= \left(1.49 \times 10^{-4}, 3.8 \times 10^{-2}, 2\times 10^{-3}, 10^{-2}, 10^{-2}\right)$.

As clearly displayed in Fig. \ref{fig1}, all four trajectories tend to converge to a non-de Sitter fixed point (the red point) as the dynamical time $\tau$ involves from zero to sixty. In addition, the de Sitter fixed point (the black point) just acts as a repeller. 
\subsection{Special limits} \label{special-limit}
Now, we would like to consider two special limits mentioned above for completeness. 

{\it (i) Second-order limit}: For this case, we do not need the introduction of $Q$, $Q_2$, $Q_3$, and $Q_4$ since the highest-order derivative is the second-order time derivative as displayed in Eqs. \eqref{limit-2nd-limit-13} and \eqref{limit-2nd-limit-14}. It is apparent that $B$ is the only dynamical variable relevant to construct the corresponding dynamical system, which turns out to be
\begin{equation} \label{dyn-1-second-order-limit}
B' =  \frac{3B \left(\alpha  B^2-144 \gamma _3-36 \gamma _4-9 \gamma _5-24 \gamma _7-4 \gamma _8\right)}{\alpha  B^2+1512 \gamma _3+432 \gamma _4+135 \gamma _5+360 \gamma _7+96 \gamma _8} ,
\end{equation}
thanks to Eq. \eqref{limit-2nd-limit-14}. It is straightforward to figure out the corresponding de Sitter  fixed point with $B\neq 0$ to this dynamical system. As a result, we have the corresponding value of $B$ given by
\begin{equation}
B^2= \frac{1}{\alpha} \left( 144 \gamma _3+ 36 \gamma _4+ 9 \gamma _5+24 \gamma _7+4 \gamma _8\right),
\end{equation}
provided $\alpha \neq 0$. This de Sitter fixed point is identical to that found in the sixth-order case. However, we can further simplify this solution as follows
\begin{equation}
B^2 = \frac{\gamma_5}{\alpha},
\end{equation}
with the help of a suitable linear combination of both constraint equations \eqref{2nd-limit-11} and \eqref{2nd-limit-12}. Since $\alpha >0$ as required in the GR, $\gamma_5$ must be positive definite. This is a unique point of this second-order limit. 

\begin{figure}[hbtp] 
 \centering
	  \includegraphics[scale=0.45]{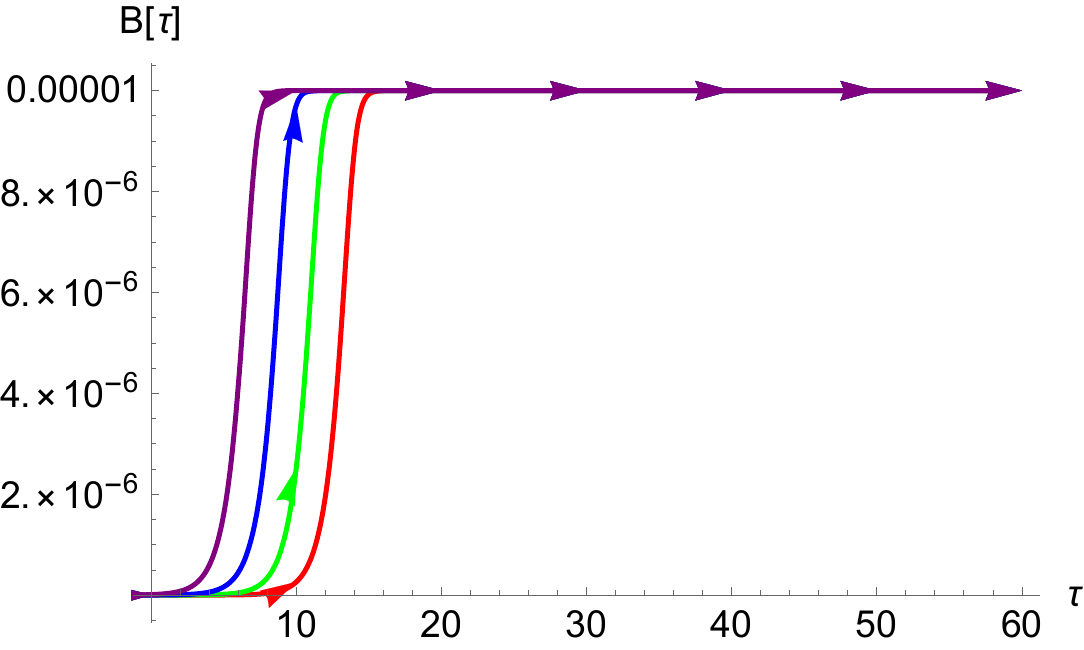}\\
\caption{\it  The de Sitter fixed point of the second-order limit acts as an attractor. The parameters have been chosen as $\alpha=1$, $\gamma_3=\gamma_4=\gamma_5=10^{-10}$, $\gamma_7=-\left(36\gamma_3+8\gamma_4 +\gamma_5 \right)/4$, $\gamma_8=\left(36\gamma_3 +6\gamma_4 -\gamma_5 \right)/2$, and  $M_p = 1$ such that the corresponding value of the de Sitter fixed point turns out to be $B=10^{-5}$. Different initial conditions are labeled by different colors as follows:  the red, green, blue, and purple curves correspond to $B[0] = 10^{-11}$, $B[0] = 10^{-10}$, $B[0] = 10^{-9}$, and $B[0] = 10^{-8}$, respectively.}
\label{fig2}
\end{figure}

Next, we are going to perturb the dynamical system formed by Eq. \eqref{dyn-1-second-order-limit} around this fixed point. It turns out that
\begin{equation}
\delta B' = \frac{6\alpha B^2 \delta B}{\alpha  B^2+1512 \gamma _3+432 \gamma _4+135 \gamma _5+360 \gamma _7+96 \gamma _8} = -3\delta B,
\end{equation}
with the help of the constraint  Eq. \eqref{2nd-limit-11}. As a result, a non-trivial solution of this first-order ODE can be easily solved to be
\begin{equation}
\delta B \propto e^{-3\tau}.
\end{equation}
This solution clearly implies an important consequence that $\delta B \to 0$ as $\tau \to +\infty$, meaning that the obtained de Sitter fixed point is apparently stable against perturbations. Furthermore, numerical calculations shown in Fig. \ref{fig2} clearly indicate that this de Sitter fixed point is an attractor of the dynamical system. Therefore, one can claim that this second-order limit is suitable for the late-time accelerated expansion of universe. 

 The obtained result in this limit provide one more vivid example supporting our observation in the recent paper \cite{cubic} that stable de Sitter solutions tend to emerge from the second-order gravity theories, whereas unstable ones seem to exist in higher-order cases such as fourth-order gravities. 
 
{\it (ii) Fourth-order limit}: For this limit, we only need the dynamical variables $B$, $Q$, and $Q_2$ to defining the corresponding dynamical system of fourth-order field equations \eqref{4th-limit-1} and \eqref{4th-limit-2} with the constraint equation \eqref{4-order-constraint}. 

As a result, the corresponding autonomous equations of this limit is given by
\begin{align}
  \label{4th-Dyn-1}
 B' &= -2QB,\\
  \label{4th-Dyn-2}
 Q' &=Q_2 -2Q^2,\\
 \label{4th-Dyn-3}
 Q_2 ' & = \frac{\beta^{(4)}}{\dot\beta^4} -3Q Q_2,
\end{align}
where $ \frac{\beta^{(4)}}{\dot\beta^4}$ can be obtained from Eq. \eqref{4th-limit-2},
\begin{align}
 \frac{\beta^{(4)}}{\dot\beta^4} =  \frac{1}{2} \frac{\Sigma_1}{ \Sigma_2},
 \end{align}
 where
 \begin{align}
  \Sigma_1=& -\alpha \left(2 Q+3 \right) B^2 -6 \left(3 \beta _1+\beta _2\right) \left(3 Q^2+6 Q+4 Q_2 \right) B \nonumber\\
  & +4 \left(60 \gamma _1+13 \gamma _2-216 \gamma _3-66 \gamma _4-24 \gamma _5-60 \gamma _7-18 \gamma _8\right) Q^3 \nonumber\\
  & -9 \left(12 \gamma _2+756 \gamma _3+216 \gamma _4+69 \gamma _5+180 \gamma _7+48 \gamma _8\right) Q^2 \nonumber\\
  &+4 \left(249 \gamma _1+61 \gamma _2-756 \gamma _3-228 \gamma _4-81 \gamma _5-204 \gamma _7-60 \gamma _8\right) Q Q_2\nonumber\\
  & -2 \left(216 \gamma _1+90 \gamma _2+1512 \gamma _3+432 \gamma _4+135 \gamma _5+360 \gamma _7+96 \gamma _8 \right) Q \nonumber\\
  &+6 \left(27 \gamma _1+7 \gamma _2-36 \gamma _3-12 \gamma _4-5 \gamma _5-12 \gamma _7-4 \gamma _8\right) Q_2^2\nonumber\\
  &+4 \left(9 \gamma _1-3 \gamma _2-648 \gamma _3-180 \gamma _4-54 \gamma _5-144 \gamma _7-36 \gamma _8\right) Q_2\nonumber\\
  &+3 \left(144 \gamma _3+36 \gamma _4+9 \gamma _5+24 \gamma _7+4 \gamma _8\right),\\
 \Sigma_2=&~ 2\left(3\beta_1+\beta_2 \right) B - \left(126 \gamma _1+36 \gamma _2-108 \gamma _3-36 \gamma _4-15 \gamma _5-36 \gamma _7-12 \gamma _8 \right)Q \nonumber\\
 & -2 \left(69 \gamma _1+22 \gamma _2-108 \gamma _3-30 \gamma _4-9 \gamma _5-24 \gamma _7-6 \gamma _8\right).
\end{align}

It is straightforward to show that the de Sitter fixed point of the sixth-order case found above, i.e.,
\begin{equation}
B^2 = \frac{1}{\alpha}\left( 144 \gamma_3 + 36 \gamma_4 + 9 \gamma_5 + 24 \gamma_7 +4\gamma_8\right), \quad Q=Q_2 =0,
\end{equation}
 is also that of this fourth-order limit. Now, our concern is about its stability. As a result, the perturbation of $\frac{\beta^{(4)}}{\dot\beta^4}$ about the de Sitter fixed point is given by
 \begin{equation} \label{4th-limit-perturbation}
 \delta \left(\frac{\beta^{(4)}}{\dot\beta^4}\right) = \frac{1}{q_4} \left(-6 \alpha B \delta B + q_5 \delta Q +q_6 \delta Q_2 \right),
 \end{equation}
 with 
 \begin{align}
 q_4 =&~ 4 \left(3\beta_1+\beta_2 \right)B - 4 \left(69 \gamma _1+22 \gamma _2-108 \gamma _3-30 \gamma _4-9 \gamma _5-24 \gamma _7-6 \gamma _8\right),\\
  q_5 =& -2\alpha B^2 -36 \left(3\beta_1+\beta_2 \right) B -2 \left(216 \gamma _1+90 \gamma _2+1512 \gamma _3+432 \gamma _4+135 \gamma _5+360 \gamma _7+96 \gamma _8 \right) ,\\
 q_6=&-24\left(3\beta_1+\beta_2 \right) B +4 \left(9 \gamma _1-3 \gamma _2-648 \gamma _3-180 \gamma _4-54 \gamma _5-144 \gamma _7-36 \gamma _8\right).
 \end{align}
 Similar to the sixth-order case, this result can be further simplified by using another perturbation equation coming from the Friedmann constraint equation \eqref{4th-limit-1},
  \begin{align}
  \alpha B \delta B +2\left[ \left(3\beta_1 +\beta_2 \right)B - 15\gamma_1 - 4\gamma_2  +108 \gamma_3 +30\gamma_4 +9 \gamma_5+24 \gamma_7+6\gamma_8  \right] \left( \delta Q_2 +3\delta Q \right) =0.
 \end{align} 
 Indeed, figuring out $ \alpha B \delta B$ from this equation and inserting it into Eq. \eqref{4th-limit-perturbation} will help us to simplify $ \delta \left(\frac{\beta^{(4)}}{\dot\beta^4}\right) $ as follows
 \begin{equation} 
  \delta \left(\frac{\beta^{(4)}}{\dot\beta^4}\right) = \frac{1}{q_4} \left(\bar q_5 \delta Q +\bar q_6 \delta Q_2 \right),
 \end{equation}
 where 
 \begin{align}
 \bar q_5 =&- 2 \left(\alpha B^2 + 486 \gamma _1+162 \gamma _2-432 \gamma _3-108 \gamma _4-27 \gamma _5-72 \gamma _7-12 \gamma _8 \right),\\
 \bar q_6 =& -12 \left[ \left(3 \beta _1 +\beta _2 \right) B+12 \gamma _1+5 \gamma _2+108 \gamma _3+30 \gamma _4+9 \gamma _5+24 \gamma _7+6 \gamma _8 \right].
 \end{align}
 
 This result will help to form a set of perturbation equations,   $\delta B' = -2 B \delta Q$,  $\delta Q' = \delta Q_2$, and $\delta Q_2 ' = \delta \left(\frac{\beta^{(4)}}{\dot\beta^4}\right)$, as a single matrix equation given by
 \begin{equation} \label{stability-equation-4th}
 \bar {\cal A}\left( {\begin{array}{*{20}c}
   A_B  \\
   A_Q  \\
   A_{Q_2} \\
 \end{array} } \right) \equiv \left[ {\begin{array}{*{20}c}
   {\lambda} & {2B} & {0 }   \\
   { 0} & {\lambda} & {-1 }  \\
     {0  } & { -\frac{\bar q_5}{q_4} } & {  \lambda-\frac{\bar q_6}{q_4}  } \\
          \end{array} } \right]  \left( {\begin{array}{*{20}c}
   A_B  \\
   A_Q  \\
   A_{Q_2} \\
 \end{array} } \right) = 0.
\end{equation}
Mathematically, this homogeneous system admits non-trivial solutions if and only if 
\begin{equation}
\det \bar {\cal A} =0,
\end{equation}
which can be expanded to the following algebraic equation of $\lambda$ defined as
\begin{equation}
\lambda \left( \bar a_2 \lambda^2+ \bar a_1 \lambda+ \bar a_0 \right) =0 ,
\end{equation}
where
\begin{align} \label{bar-a2}
 \bar a_2 =&~ 2 \left[ \left(3 \beta _1 +\beta _2\right) B+ \gamma _2+108 \gamma _3+30 \gamma _4+9 \gamma _5+24 \gamma _7+6 \gamma _8\right],\\
 \bar a_1 =&~ 3 \bar a_2,\\
 \bar a_0 =& -2 \alpha  B^2 <0.
\end{align}
It now becomes clear that the positivity of $\bar a_2$ will lead to the existence of at least one positive root $\lambda>0$ of the above equation, besides the trivial one $\lambda=0$, since $\bar a_0 <0$.  Therefore, the corresponding fourth-order limit may be suitable for the inflationary phase of early universe. Otherwise, the fourth-order limit may be suitable for the accelerated expansion of late-time universe.

\begin{figure}[hbtp] 
 \centering
	  \includegraphics[scale=0.5]{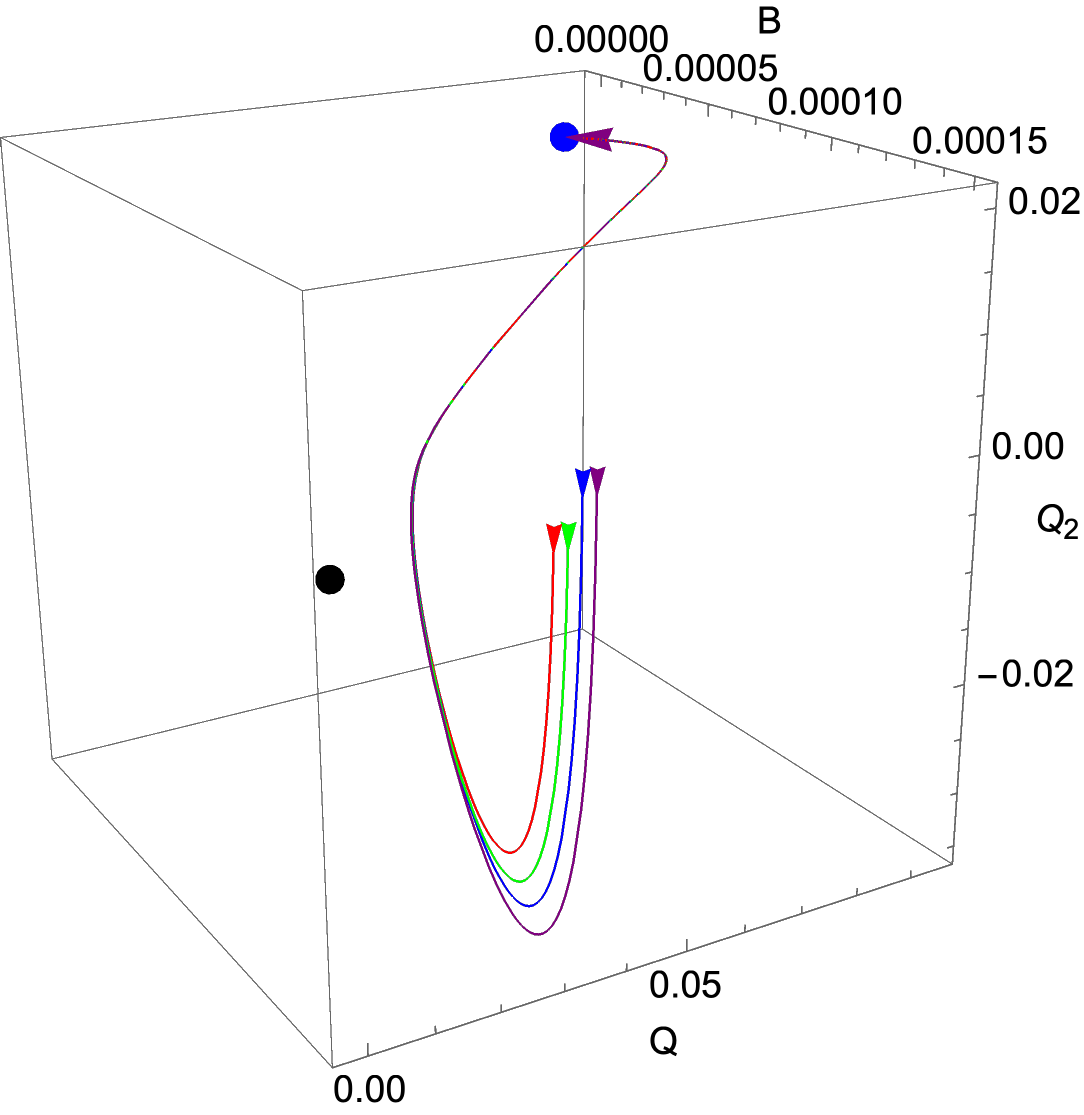}\\
\caption{\it The de Sitter fixed point  of the fourth-order limit with $\left(B,Q,Q_2\right)\simeq \left(1.473\times 10^{-4},0,0\right)$, displayed as a black point, acts as an repeller. An attractor with $\left(B,Q,Q_2\right) \simeq \left(0,9.076 \times 10^{-2}, 1.648 \times 10^{-2} \right)$, displayed as a blue point, corresponds to a non-de Sitter fixed point associated with a real solution $Q$ of Eq. \eqref{non-de-Sitter-fixed-point-4th-limit}. Initial conditions for the trajectories are identical to that used for the case of the sixth-order gravity.}
\label{fig3}
\end{figure}

To have a complete view, it is useful to define other fixed points, which will be called non-de Sitter fixed points and correspond to
\begin{equation}
B=0,\quad Q_2 =2Q^2,\quad \frac{\beta^{(4)}}{\dot\beta^4} =3Q Q_2.
\end{equation}
Consequently, we are able to define the corresponding equation of $Q$ from the field equations such as
\begin{equation} \label{non-de-Sitter-fixed-point-4th-limit}
\left(2Q+1 \right) \left( \bar\kappa_3 Q^3+\bar \kappa_2 Q^2+ \bar\kappa_1 Q+\bar\kappa_0 \right)=0,
\end{equation}
where the coefficients $\bar\kappa_i$ $(i=0-4)$ are given by
\begin{align}
\bar\kappa_3=& ~10 \left(6 \gamma _1-36 \gamma _3-12 \gamma _4-5 \gamma _5-12 \gamma _7-4 \gamma _8\right),\\
\bar\kappa_2=&~ 3 \left(28 \gamma _1-468 \gamma _3-136 \gamma _4-45 \gamma _5-116 \gamma _7-32 \gamma _8\right),\\
\bar\kappa_1 =&~36 \left(\gamma _1-36 \gamma _3-10 \gamma _4-3 \gamma _5-8 \gamma _7-2 \gamma _8\right),\\
\bar\kappa_0 =&~144 \gamma _3+36 \gamma _4+9 \gamma _5+24 \gamma _7+4 \gamma _8.
\end{align}
It should be noted that the constraint \eqref{4-order-constraint} has been used to simplify this equation. Similar to the sixth-order case, this equation always admits a parameter-independent solution $Q=-1/2$ besides three parameter-dependent ones, whose value will be determined by parameters $\gamma_i~ (i=1, ~3, ~4,~5,~7,~8)$. Interestingly, one of the non-de Sitter fixed points will act as an attractor to the dynamical system as shown in Fig. \ref{fig3}. 

 \subsection{Effect of the quadratic curvature terms}
In this subsection, we would like to see  the effect of the quadratic curvature terms, i.e., $R^2$ and $R_{\mu\nu}^2$, on the stability of the obtained de Sitter solutions. In particular, these terms will be turned off, meaning that $\beta_1$ and $\beta_2$ will be set to be zero. It is interesting to note that both these quadratic curvature terms generate the similar derivative terms in the field equations  \eqref{first-field-equation} and \eqref{second-field-equation}. This implies that they share the same effect on the stability of the obtained de Sitter solutions. For the sixth-order gravity case, it turns out that $\beta^{(6)}$ does not couple to any other derivatives in the field equations \eqref{first-field-equation} and \eqref{second-field-equation}. Therefore, we can always define the quantity $\frac{\beta^{(6)}}{\dot\beta^6}$  along with its perturbation  $\delta \left(\frac{\beta^{(6)}}{\dot\beta^6} \right)$, in contrast to $\frac{\beta^{(4)}}{\dot\beta^4}$ in the generalized Einsteinian cubic gravity \cite{cubic}. This result indicates that we can always define a complete set of perturbed equations for the sixth-order gravity, even in the absence of the quadratic curvature terms. 
Now, we turn to the fourth-order limit, in which $\beta^{(6)}$ and $\beta^{(5)}$ no longer exist. To be more specific, see Eq. \eqref{4th-limit-2}.  It appears in the $\gamma_j$ terms that although $\beta^{(4)}$ couples to $\ddot\beta$ but it also couples to $\dot\beta^2$. Hence, we can still define explicitly $\frac{\beta^{(4)}}{\dot\beta^4}$  along with its perturbation  $\delta \left(\frac{\beta^{(4)}}{\dot\beta^4} \right)$ as long as $\Sigma_2 \neq 0$ and $q_4 \neq 0$, even when the quadratic curvature terms, $R^2$ and $R_{\mu\nu}^2$, disappear.  This claim can be easily verified by the analysis presented in subsection \ref{special-limit}, especially Eq. \eqref{4th-limit-perturbation}. This interesting result indicates that the quadratic curvature terms, e.g., $R^2$, could only be important in quadratic and cubic curvature gravities such as that proposed in Ref. \cite{DeFelice:2023vmj}. In addition, fourth-order limit(s) of higher-than-cubic-order curvature gravities may not need the help of the quadratic curvature terms, e.g., $R^2$, in order to have a complete set of perturbed equations \cite{cubic}. 
\section{Conclusions}\label{final}
Inspired by our previous studies \cite{cubic,Do:2020vdc,Do:2021fal,Do:2023yvg,Pham:2024fub,Do:2026pkg,preprint}, we would like to examine whether the sixth-order gravity \cite{Giacchini:2025gzw,Giacchini:2024exc} admits a stable de Sitter solution as its cosmological solution. As a result, we have successfully derived the field equations to the model using the effective Euler-Lagrange equation method. Then, the field equations have been solved analytically to give us the exact de Sitter solution. To investigate whether this solution is stable or not, the corresponding dynamical system of the sixth-order gravity has been constructed from the field equations. As expected, this system admits a de Sitter fixed point, which is exactly equivalent to the de Sitter solution, besides some non-de Sitter fixed points. As a result, the value of the obtained de Sitter solution is solely defined by all cubic curvature terms.  Interestingly, it turns out that we do not meet any problem in defining the set of perturbed equations from this system even when the quadratic curvature terms are absent,  in contrast to the generalized Einsteinian cubic gravity \cite{cubic}. This result provides a hint that the quadratic curvature terms, $R^2$ and $R_{\mu\nu}^2$, could only matter in quadratic and cubic gravities in particular or fourth-order gravities in general. 

More interestingly, although the coefficients $\gamma_1$ and $\gamma_2$ of the terms $R\Box R$ and $R_{\mu\nu} \Box R^{\mu\nu}$, respectively, do not contribute to the value of the obtained de Sitter solution, but they do affect on the stability of this homogeneous and isotropic solution. In particular,  we have shown that the de Sitter solution of the sixth-order gravity will be unstable if $3\gamma_1 +\gamma_2 <0$. This result indicates that two quadratic curvature terms, $R^2$ and $R_{\mu\nu}R^{\mu\nu}$, are less important in terms of stability. However, these quadratic curvature terms will play a non-trivial role in the fourth-order limit, in which $3\gamma_1 +\gamma_2 =0$ such that all sixth- and fifth-order derivatives disappear. In particular, it has been shown that the de Sitter solution in this limit will be unstable once $\bar a_2$ shown in Eq. \eqref{bar-a2} is positive definite. Furthermore, the sign of $\bar a_2$ is partially determined by the coefficients of $R^2$ and $R_{\mu\nu}R^{\mu\nu}$, i.e., $\beta_1$ and $\beta_2$, respectively.  As a result, one could conclude, in the light of discussions in Refs. \cite{Elizalde:2014xva,Pozdeeva:2019agu,Vernov:2021hxo}, that these scenarios seem to be more suitable for the inflationary phase of early universe because they would not face  the so-called eternal inflation and related multiverse issue \cite{Guth:2007ng,Brustein:1994kw}. For people seeking a stable de Sitter solution, the second-order limit may be relevant, provided the constraints in Eqs. \eqref{2nd-limit-1}, \eqref{2nd-limit-10}, \eqref{2nd-limit-11}, and \eqref{2nd-limit-12} are all satisfied.

According to the obtained results, the sixth-order gravity model described by the action in Eq. \eqref{action-1} along with its fourth-order limit would be promising extensions of the Starobinsky model for consistency with the recent ACT data. It would be interesting to extend works done in Ref. \cite{Khodabakhshi:2024med} to the sixth-order gravity to see if it is cosmologically viable.  Besides, it is possible to apply our current analysis to higher-than-sixth-order gravities. An interesting example is an eighth-order gravity involving $R \Box^2 R$ considered in Ref. \cite{BattagliaMayer:1993yf}. We will leave these interesting issues to our future works.
\begin{acknowledgments}
This study is funded by the Vietnam National Foundation for Science and Technology Development (NAFOSTED) under grant number 103.01-2023.50. The author would also like to thank Prof. Phung V. Dong very much for his supports. 
\end{acknowledgments}
\appendix
\section{Cross-check} \label{app}
In this Appendix, we would like to perform a cross-check, by which one can easily verify the validity of our stability analysis. This work follows our previous one done in Ref. \cite{cubic} as well as other work by other people in Ref. \cite{CamposDelgado:2024jst}. As a result, this approach is not based on the dynamical system. It follows a direct perturbation of metric fields around the de Sitter solution. Additionally, it only uses the $00$-component of the Einstein equation shown in Eq. \eqref{first-field-equation}. In particular, by taking a perturbation of $\beta(t)$ around the de Sitter solution with $\dot\beta = \zeta \neq 0$ and $\beta^{(n \geq 2)}=0$, 
\begin{equation}
\beta(t) \to \beta(t) +\delta \beta(t),
\end{equation}
Eq. \eqref{first-field-equation} will be perturbed as follows
\begin{align}
&2\alpha \dot\beta \delta \dot\beta +2 \left(3\beta_1 +\beta_2 \right) \left(2\dot\beta \delta \beta^{(3)} +6 \dot\beta^2 \delta \ddot\beta \right) -6\gamma_1 \left(2\dot\beta \delta \beta^{(5)} +12 \dot\beta^2 \delta \beta^{(4)} +10 \dot\beta^3 \delta \beta^{(3)} -24 \dot\beta^4 \delta \ddot\beta  \right) \nonumber\\
&-2\gamma_2 \left(2\dot\beta \delta \beta^{(5)} +12 \dot\beta^2 \delta \beta^{(4)} +8 \dot\beta^3 \delta \beta^{(3)} -30 \dot\beta^4 \delta \ddot\beta  \right) -36\gamma_3 \left[2\dot\beta^4 \left(\delta \ddot\beta +4\dot\beta \delta \dot\beta \right) -2 \dot\beta^2 \left( 6\dot\beta \delta \beta^{(3)} +19 \dot\beta^2 \delta \ddot\beta -8\dot\beta^3 \delta \dot\beta \right) \right] \nonumber\\
&+12\gamma_4 \left(10\dot\beta^3 \delta \beta^{(3)} +30 \dot\beta^4 \delta \ddot\beta -18\dot\beta^5 \delta \dot\beta \right) + \gamma_5 \left(36\dot\beta^3 \delta \beta^{(3)} +108 \dot\beta^4 \delta \ddot\beta -54\dot\beta^5 \delta \dot\beta \right) \nonumber\\
&+12\gamma_7 \left(8\dot\beta^3 \delta \beta^{(3)} +24 \dot\beta^4 \delta \ddot\beta -12\dot\beta^5 \delta \dot\beta \right) + 4\gamma_8 \left(6\dot\beta^3 \delta \beta^{(3)} +18 \dot\beta^4 \delta \ddot\beta -6\dot\beta^5 \delta \dot\beta \right) =0.
\end{align}
Similar to Ref. \cite{cubic}, we consider an exponential perturbation such as
\begin{equation}
\delta \beta (t) = C_\beta e^{\hat\lambda \zeta t},
\end{equation}
which implies the following results,
\begin{equation}
\delta \dot\beta = \hat\lambda \zeta \delta \beta,~\delta \beta^{(n\geq 2)} = \hat\lambda^n \zeta^n \delta \beta.
\end{equation}
As a result, we are able to obtain the corresponding polynomial equation of $\hat\lambda$ given by
\begin{equation} \label{poly-eq}
\hat\lambda \left( c_4 \hat\lambda^4+c_3 \hat\lambda^3+c_2 \hat\lambda^2+c_1 \hat\lambda^1+c_0 \right)=0,
\end{equation}
where
\begin{align}
c_4&= -2\zeta ^4 \left(3 \gamma _1 +\gamma_2 \right) ,\\
c_0& =\alpha - 3 \zeta^4 \left(144 \gamma _3 +36 \gamma _4 +9 \gamma _5 +24 \gamma _7+4  \gamma _8 \right).
\end{align}
Here, we do not list other coefficients since they are not important for further discussions. It appears that $c_0$ can be reduced, thanks to the de Sitter solution shown in Eq. \eqref{value-of-zeta}, to a simple value given by
\begin{equation}
c_0 = -2\alpha <0.
\end{equation}
It now becomes clear that if $3\gamma_1+\gamma_2 <0$ then $c_4>0$. Consequently, Eq. \eqref{poly-eq} will admit at least one positive root $\hat\lambda >0$ besides the trivial one $\hat\lambda =0$ since $c_0 c_4 <0$. This result is really consistent with our analysis obtained in Subsection \ref{stability}. For now, we  conclude that the stability analysis based on the dynamical system method considered in the main text is indeed valid. 

\end{document}